\documentclass[
  aps,
  reprint,
  prb,
  showpacs,
  superscriptaddress,
  longbibliography,
]{revtex4-1}
\pdfoutput=1

\usepackage{graphicx}
\usepackage{amsmath}
\usepackage{amssymb}
\usepackage{amsfonts}
\usepackage{color}
\usepackage{hyperref}
\usepackage{bm}

\newcommand{\psireal}{\psi}
\newcommand{\psichiral}{\tilde{\psi}}

\newcommand{\ptitle}{Bosonic Gaussian states from conformal field theory}
\newcommand{\MPQ}{Max-Planck-Institut f{\"u}r Quantenoptik, Hans-Kopfermann-Stra{\ss}e\ 1, D-85748 Garching, Germany}
\newcommand{\kwds}{strongly correlated systems; Gaussian states; conformal field theory}

\hypersetup{
  pdftitle={\ptitle},
  pdfkeywords={\kwds},
  pdfauthor={Benedikt Herwerth, Germ{\'a}n Sierra, J. Ignacio Cirac, Anne E. B. Nielsen},
  pdfsubject={},
  pdfcreator={pdflatex},
  colorlinks,
  linkcolor=blue,
  citecolor=blue,
  urlcolor=blue
}

\begin{document}
\author{Benedikt Herwerth}
\email{benedikt.herwerth@mpq.mpg.de}
\affiliation{\MPQ}
\author{Germ\'an Sierra}
\affiliation{Instituto de F{\'i}sica Te{\'o}rica, UAM-CSIC, Madrid, Spain}
\author{J. Ignacio Cirac}
\affiliation{\MPQ}
\author{Anne E. B. Nielsen}
\altaffiliation{On leave from: Department of Physics and Astronomy, Aarhus University, DK-8000 Aarhus C, Denmark}
\affiliation{Max-Planck-Institut f{\"u}r Physik komplexer Systeme, D-01187 Dresden, Germany}
\title{\ptitle}

\begin{abstract}
We study nonchiral wave functions for systems with continuous spins
obtained from the conformal field
theory (CFT) of a free, massless boson. In contrast to the
case of discrete spins, these can be treated as bosonic Gaussian states,
which allows us to efficiently
compute correlations and entanglement properties both in
one (1D) and in two spatial dimensions (2D). In 1D, the computed
entanglement entropy and spectra are in agreement with the underlying
CFT. Furthermore, we construct a 1D parent Hamiltonian with a low-energy
spectrum corresponding to that of a free, massless boson.
In 2D, we find edge excitations in the entanglement spectrum,
although the states do not have intrinsic topological order,
as revealed by a determination of the topological entanglement
entropy.
\end{abstract}

\maketitle

\section{Introduction}
The main challenge in the theoretical study of many-body
systems is given by the exponential scaling of the Hilbert
space with the system size. Since exact diagonalization
techniques are limited to small systems, studying simple
models is essential for understanding intricate many-body phenomena.
One way to build such models is to define them through Hamiltonians.
A different approach is based on wave functions
representing model states in a variational sense. The most prominent
example is given by Laughlin's wave
function~\cite{Laughlin1983} and
its generalizations~\cite{Haldane1983, Halperin1984, Moore1991}. 
These provide good variational descriptions of electrons
in a fractional quantum Hall~\cite{Tsui1982} (FQH) phase
and thus represent paradigmatic models of systems
exhibiting topological order~\cite{Wen1990b, Wen2016}.
Laughlin's wave function is the two-dimensional (2D) analog
of the ground state belonging
to the one-dimensional (1D) Calogero-Sutherland
model~\cite{Calogero1969, Calogero1971, Sutherland1971, Sutherland1971a}.

In the past years, there has been an increased interest
in lattice models exhibiting topological properties
and in implementing these with systems
consisting of ultracold
atoms, see for example
Refs.~\onlinecite{Sorensen2005, Hafezi2007, Cooper2013, Yao2013, Goldman2016}.
An advantage of this approach is that the
interaction strength
can be tuned experimentally~\cite{Bloch2008}.
Thus, strongly interacting regimes can be achieved where extraordinary
many-body effects are expected to be more pronounced and
occur at higher temperatures.
Recent experimental achievements in this field include the
implementation of the Hofstadter model~\cite{Aidelsburger2013, Miyake2013}
and the simulation of a four-dimensional
FQH effect~\cite{Lohse2018, Zilberberg2018}.
Important theoretical models on lattices
are given by the chiral spin liquid~\cite{Kalmeyer1987, Kalmeyer1989},
which is the lattice version of Laughlin’s wave function,
its nonabelian generalization~\cite{Greiter2009},
and also the Haldane-Shastry model~\cite{Haldane1988, Shastry1987},
which is the lattice analog of the Calogero-Sutherland model.
Recently, an exact parent Hamiltonian
of the chiral spin liquid
was found~\cite{Thomale2009, Nielsen2012},
which led to a proposal for an implementation with
ultracold atoms~\cite{Nielsen2013a}.

In a seminal paper, Moore and Read~\cite{Moore1991} showed that
Laughlin's and other FQH wave functions
can be constructed systematically using
conformal field theory (CFT)
in 1(+1) dimension. In this approach, a FQH wave function
is defined as a chiral CFT correlator.
Since CFT is associated with the
gapless boundary
excitations~\cite{Halperin1982, Wen1990c, Wen1990d, Wen1991, Wen1992}
of a quantum Hall system, this is an
example of a bulk-edge correspondence~\cite{Dubail2012a}.
The description of a FQH state
in terms of CFT thus establishes a connection between the topological
order of the state and its edge theory~\cite{Witten1989, Dubail2012a}.
The idea of building model states
from CFT was later also applied to lattice systems
with discrete spin or fermionic degrees of freedom~\cite{Cirac2010}.
In 1D, ground
states of Haldane-Shastry spin chains were obtained in this
way~\cite{Cirac2010, Nielsen2011}, while
these states describe quantum Hall
lattice systems in 2D.
A wave function of
free fermions occurs as a
special case having a vanishing topological
entanglement entropy~\cite{Nielsen2012}.
It is an example of a chiral state with non-intrinsic
topological order and represents a lattice
version of the integer quantum Hall~\cite{Klitzing1980} effect.
Other 2D lattice states obtained from CFT were shown
to exhibit intrinsic topological order
and describe FQH phases~\cite{Nielsen2012, Tu2013, Glasser2016}.

Given the construction of model states in terms of CFT, it is natural
to ask how their physical properties are related to the CFT they are
constructed from. For states in 1D, it was shown that correlations and
entanglement entropies are in accordance with the CFT
expectation~\cite{Tu2013}.
In 2D, CFT is associated with the gapless edge of FQH systems.
As shown in Ref.~\onlinecite{Dubail2012a} for
the case FQH states with continuous spatial degrees of freedom,
the assumption of exponentially decaying bulk correlations
implies that the edge correlations are determined by CFT.
Except in special cases
like that of free fermions~\cite{Nielsen2012},
the determination of correlations and entanglement entropies
of states defined through CFT requires the use of Monte Carlo for large
systems, where exact methods fail. Recently~\cite{Herwerth2017},
we made an approximation
for correlations in a class of
abelian FQH lattice states and the corresponding 1D wave functions.
This approximation corresponds
to replacing the discrete spin-$\frac{1}{2}$ degrees of
freedom by continuous spins, which makes it possible to obtain exact
results for large systems. We found good quantitative agreement
between the actual correlations and the approximation
in a certain parameter range.
In 2D, the approximation has polynomially decaying edge
correlations and exponentially decaying bulk correlations.

In this paper, we adopt Moore and Read's approach and define
a class of spin states on lattices from the CFT of a massless, free boson.
Opposed to earlier studies on lattices, we consider the case of continuous
spins $s \in \mathbb{R}$. On the one hand, this implies that these states
have the correlations that we used earlier~\cite{Herwerth2017} to
approximate the spin-$\frac{1}{2}$ case. On the other hand, they represent
another example of model states constructed from CFT,
and the aim of this paper is to investigate and characterize them.
The main motivation for studying the case of continuous spins
is that the resulting wave functions are Gaussian.
Opposed to the discrete case, their properties can thus
be computed efficiently using the framework of bosonic Gaussian states.
Following a new development of the past
years~\cite{Vidal2003, Calabrese2004, Kitaev2006, Levin2006, Li2008},
we use entanglement properties as the central tool
to characterize the states and investigate their topological
properties.

In 1D, the states are closely related to the CFT they are
constructed from: We show that their entanglement
entropies and low-lying momentum-space entanglement energies
agree with those of a massless, free boson.
We also construct a parent Hamiltonian in 1D with
low-lying energies corresponding
to that of the CFT.
In 2D, we confirm that the states do not have intrinsic
topological order since
their topological entanglement entropy~\cite{Levin2006, Kitaev2006}
vanishes as expected for Gaussian states.
On the other hand, we observe low-lying states in the entanglement spectrum
that are exponentially localized at the edge.
The wave functions studied in this paper are constructed
from the complete (chiral + antichiral) CFT
and are thus real and nonchiral. This absence
of chirality is similar to the
quantum spin Hall effect~\cite{Kane2005, Kane2005b, Bernevig2005}.
We also comment on the corresponding
chiral wave function, which is equivalent to the real
case in 1D. For general lattices, however, the chiral state
depends on the ordering of
the lattice positions, which is in contrast to the real wave function.
This could indicate that the chiral case
does not fall under the framework of bosonic Gaussian states, and
we leave this case open for future investigations.

This paper is structured as follows:
Sec.~\ref{sec:Gaussian-states-definitions}
defines the states with continuous spins studied in this paper,
shows how they can be represented as bosonic Gaussian states,
and explains how we compute entanglement properties for them.
We discuss results for a 1D system with periodic boundary conditions in Sec.~\ref{sec:properties-of-states-in-1D} and for 2D systems on the
cylinder in Sec.~\ref{sec:properties-of-states-in-2D}.
Sec.~\ref{sec:conclusion} concludes this paper.

\section{Spin states from conformal field theory}
\label{sec:Gaussian-states-definitions}
This section defines states with a continuous spin
as correlators of the free-boson CFT,
represents them as bosonic Gaussian states,
and explains how we compute entanglement properties.

\subsection{Definition of states}
We consider the CFT of a massless, free bosonic field
$\varphi(z, \bar{z})$ for $z \in \mathbb{C}$.
This theory has a series of conformal
primary fields $:e^{i s \varphi(z, \bar{z})}:$ for $s \in \mathbb{R}$,
where the colons denote normal ordering.
We define spin wave functions as the correlator of primary fields:
\begin{align}
\label{eq:definition-psi-real-as-correlator}
&\psireal_{\beta}(\bm{s}) \\
&\quad=  e^{-\frac{1}{4} \left(\beta + \beta_{0}\right) \bm{s}^2}  \langle :e^{\frac{i}{\sqrt{2}} s_1 \varphi(z_1, \bar{z}_1)}: \dots :e^{\frac{i}{\sqrt{2}} s_N \varphi(z_N, \bar{z}_N)}: \rangle \notag \\
&\quad= e^{-\frac{1}{4} \left(\beta + \beta_{0}\right) \bm{s}^2} \delta\left(s_1 + \dots + s_N\right) \prod_{m < n} \left|z_m - z_n\right|^{s_m s_n},\notag 
\end{align}
where $z_j$ for $j \in \{1, \dots, N\}$ defines a lattice
of positions in the complex plane,
$\bm{s} \in \mathbb{R}^N$ is a vector of $N$ continuous spin
variables, $\delta$ is the Dirac delta function originating
from the charge neutrality condition,
and $\beta > 0$ is a real parameter.
We define the real number $\beta_0$ through
a normalizability criterion and introduce
$\beta_0$ separately from $\beta$ for convenience so that $\beta$
is always positive, cf.
Sec.~\ref{sec:determination-of-beta-0} below.
The parametric dependence of $\psireal_{\beta}(\bm{s})$
on the lattice positions $z_j$ is suppressed
for simplicity of notation.

The prefactor $e^{-\frac{1}{4} \left(\beta + \beta_{0}\right) \bm{s}^2}$
in Eq.~\eqref{eq:definition-psi-real-as-correlator}
corresponds to a rescaling
$(z_j, \bar{z}_j) \to (\lambda z_j, \lambda \bar{z}_j)$ with $\lambda > 0$,
under which the correlation function
of primary fields
$:e^{\frac{i}{\sqrt{2}} s_j \varphi(z_j, \bar{z}_j)}:$ changes
by a factor $\lambda^{-\frac{1}{2} \bm{s}^2}$.  Comparing to
the form of the wave function of
Eq.~\eqref{eq:definition-psi-real-as-correlator}, 
we have $\beta + \beta_0 = 2\ln \lambda$ in terms of the
scale parameter $\lambda$ of the lattice.
In the definition of $\psireal_{\beta}(\bm{s})$,
we do not include an additional parameter in the
exponent of the vertex operators as in the case of discrete
spins~\cite{Cirac2010}. The reason is that such a parameter can be removed by a rescaling of the continuous spins $s_j \in \mathbb{R}$.

In a previous study~\cite{Herwerth2017}, we considered
continuous-spin approximations for correlations of spin-$\frac{1}{2}$
states obtained from CFT. The wave functions $\psireal_{\beta}(\bm{s})$
have the same correlations as the approximation that was
made in Ref.~\onlinecite{Herwerth2017}, cf. Appendix~\ref{appendix:relation-to-previous-approximation}
for details.

In this work, we focus on the case of a real wave function
$\psireal_{\beta}(\bm{s})$ defined through operators $:e^{\frac{i}{\sqrt{2}} s_j \varphi(z_j, \bar{z}_j)}:$
with chiral and antichiral components.
We comment on the corresponding chiral wave function
in Appendix~\ref{sec:psi-cont-chiral-state},
where we also show that $\psireal_{\beta}(\bm{s})$
is equivalent to the chiral case for a uniform 1D lattice
with periodic boundary conditions.

\subsection{Representation as a Gaussian state}
\label{sec:regularization-of-delta-function}
We now represent $\psireal_{\beta}(\bm{s})$
as a bosonic Gaussian state, which implies
that its properties can be computed efficiently for large systems.
To this end, we
replace the delta function $\delta(s_1 + \dots + s_N)$ in $\psireal_{\beta}(\bm{s})$ by
a Gaussian of width proportional to $\sqrt{\epsilon}$ for $\epsilon > 0$.
This leads to the wave function
\begin{align}
\label{eq:regularized-gaussian-wave-function}
\psireal_{\beta, \epsilon}(\bm{s}) &= e^{-\frac{1}{2} \bm{s}^t (\frac{1}{2 \epsilon} \bm{e} \bm{e}^t + X_{\beta}) \bm{s}},
\intertext{where}
\label{eq:def-of-X-matrix}
\left(X_{\beta}\right)_{mn} &= \frac{1}{2} \left(\beta + \beta_0\right) \delta_{m n} + X_{mn},\\
X_{mn} &= -\ln \left(|z_m - z_n| + \delta_{m n}\right),
\end{align}
and $\bm{e} = (1, \dots, 1)^t$ is the vector with all entries being equal to one.
Then,
\begin{align}
\psireal_{\beta}(\bm{s}) &= \lim_{\epsilon \to 0} \frac{1}{2 \sqrt{\pi \epsilon}} \psireal_{\beta, \epsilon}(\bm{s}).
\end{align}
Defining
\begin{align}
\label{eq:definition-of-X-beta-epsilon}
X_{\beta, \epsilon} &= \frac{1}{2 \epsilon} \bm{e} \bm{e}^t + X_{\beta}
\equiv \frac{1}{2 \epsilon} \bm{e} \bm{e}^t +  \frac{1}{2} \left(\beta + \beta_0\right) \mathbb{I} + X,
\end{align}
where $\mathbb{I}$ is the $N \times N$ identity matrix,
$\psireal_{\beta, \epsilon}(\bm{s})$ assumes the standard form of a
pure, bosonic Gaussian state (cf. Appendix~\ref{sec:appendix-general-Gaussian-states}):
\begin{align}
\psireal_{\beta, \epsilon}(\bm{s}) &= e^{-\frac{1}{2} \bm{s}^t X_{\beta, \epsilon} \bm{s}}.
\end{align}

\subsection{Definition of $\beta_0$}
\label{sec:determination-of-beta-0}
We define $\beta_0$ by requiring that $\psireal_{\beta, \epsilon}(\bm{s})$ is
normalizable for all $\beta > 0$.
According to Eq.~\eqref{eq:def-of-X-matrix},
we have
\begin{align}
\beta_0 = -2 \min\{\lambda_{\epsilon}^{(1)}, \dots, \lambda_{\epsilon}^{(N)}\},
\end{align}
where $\{\lambda_{\epsilon}^{(1)}, \dots, \lambda_{\epsilon}^{(N)}\}$ are the
eigenvalues of $\frac{1}{2 \epsilon} \bm{e} \bm{e}^t + X$.

In the limit $\epsilon \to 0$, the matrix $\frac{1}{2 \epsilon} \bm{e} \bm{e}^t + X$
becomes divergent, and we determine $\{\lambda_{\epsilon}^{(1)}, \dots, \lambda_{\epsilon}^{(N)}\}$ as the inverse
eigenvalues of
\begin{align}
\left[\frac{1}{2 \epsilon} \bm{e} \bm{e}^t + X\right]^{-1} &= X^{-1} - \frac{ X^{-1} \bm{e} \bm{e}^{t} X^{-1}}{2 \epsilon + \bm{e}^t X^{-1} \bm{e}},
\end{align}
where we used a general formula for the inverse of a matrix that is changed by
a term of rank one~\cite{Bartlett1951}.

\subsection{Entanglement properties}
The representation of $\psireal_{\beta}(\bm{s})$
as a Gaussian state allows us to efficiently compute
its entanglement properties under partition
of the system into parts $A$ and $B$,
cf. Appendix~\ref{sec:appendix-entanglement-properties-of-Gaussian-states-from-CFT} for details.

In summary, we find that the R\'{e}nyi entanglement entropies
$S_a(A)$ of order $a$ are given by
\begin{align}
\label{eq:Sa-and-finite-Sap}
S_a(A) &= -\frac{1}{2} \ln \epsilon + S_a'(A) + \mathcal{O}(\epsilon),
\end{align}
where $S_a'(A)$ is independent of $\epsilon$. The divergence
in $S_a(A)$ for $\epsilon \to 0$ is a consequence of
the delta function $\delta(s_1 + \dots + s_N)$ in $\psireal_{\beta}(\bm{s})$. By subtracting it, we obtain
the finite entropies $S_a'(A)$.
The entanglement Hamiltonian can be brought into the
diagonal form $\sum_{j=1}^{|A|} \tilde{\omega}_j b^\dagger_j b_j$
in a suitable basis of annihilation and creation operators $b_j$ and $b_j^\dagger$. The single-particle energies $\tilde{\omega}_j$
determine the entanglement spectrum.

\subsection{States on the cylinder}
\label{sec:cont-spin-states-cylinder}
For the rest of this paper, we study a system on a cylinder
with a square lattice of $N_x$ sites in the open and $N_y$ sites
in the periodic direction:
\begin{align}
\label{eq:psi-cont-coordinates-cylinder}
w_m &\equiv w_{m_x m_y} = \frac{2 \pi}{N_y} \left(m_x + i m_y\right),
\end{align}
where $m_x$ is the $x$ and $m_y$ the $y$ component
of the index $m$ [$m = (m_x - 1) N_y + m_y$],
and we identify $w_{m_x m_y}$ with $w_{m_x, m_y+N_y}$.
This includes a uniform lattice
in 1D with periodic boundary conditions as the special case $N_x = 1$. 

The wave function $\psireal_{\beta}(\bm{s})$ was defined for positions
$z_j$ in the complex plane in Eq.~\eqref{eq:definition-psi-real-as-correlator}. For positions $w_j$
on the cylinder, we use the map $z_j = e^{w_j}$ from the cylinder to the plane and define the wave function by
evaluating the CFT correlator on the cylinder:
\begin{align}
\label{eq:definition-psi-real-as-correlator-cylinder}
&\psireal_{\beta}(\bm{s}) \\
&\quad=  e^{-\frac{1}{4} \left(\beta + \beta_{0}\right) \bm{s}^2}  \langle :e^{\frac{i}{\sqrt{2}} s_1 \varphi(w_1, \bar{w}_1)}: \dots :e^{\frac{i}{\sqrt{2}} s_N \varphi(w_N, \bar{w}_N)}: \rangle \notag \\
&\quad= e^{-\frac{1}{4} \left(\beta + \beta_{0}\right) \bm{s}^2} \delta\left(s_1 + \dots + s_N\right) \notag \\
&\quad\quad\times \prod_{m < n} \left|2 \sinh\left(\frac{1}{2} (w_m - w_n)\right)\right|^{s_m s_n},\notag
\end{align}
where we used Eq.~\eqref{eq:definition-psi-real-as-correlator} and the transformation rule
for primary fields~\cite{DiFrancesco1997} under $z_j = e^{w_j}$.
The definition of
$\psireal_{\beta, \epsilon}(\bm{s})$ introduced in Sec.~\ref{sec:regularization-of-delta-function} changes accordingly on the cylinder.

\section{Properties of states in 1D}
\label{sec:properties-of-states-in-1D}
In this section, we study properties of $\psireal_{\beta}(\bm{s})$ for
a 1D system with periodic boundary conditions.
We show that the correlations, entanglement
properties, and a parent Hamiltonian exhibit signatures of the underlying
CFT of a free, massless boson.

\subsection{Correlations}
\begin{figure*}[htb]
\centering
\includegraphics{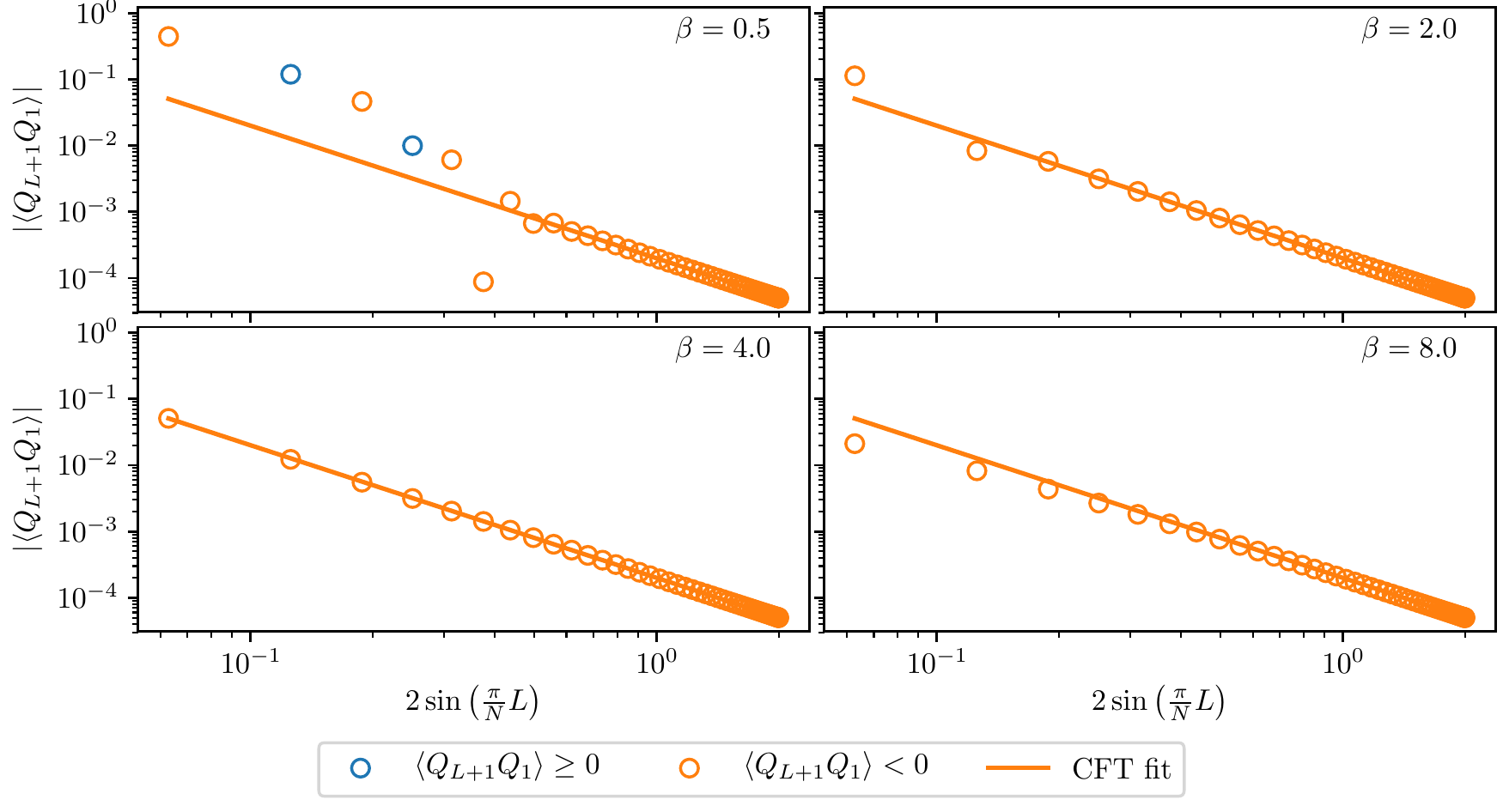}
\caption{(Color online) Spin-Spin correlations $\langle Q_{L+1} Q_1 \rangle$ in $\psireal_{\beta}(\bm{s})$ in 1D
with periodic boundary conditions for $N=100$ sites.
The operator $Q_m$
is defined as $(Q_m \psireal_{\beta})(\bm{s}) = s_m \psireal_{\beta}(\bm{s})$. The long-range decay
is consistent with the CFT expectation of an algebraic decay with
a power of $-2$. The fit was done for the 10 data points
with the largest value of $L$.
}
\label{fig:correlations-1d-in-state-psi-beta}
\end{figure*}
A plot of spin-spin correlations in $\psireal_{\beta}(\bm{s})$
and a fit to the CFT expectation are shown in
Fig.~\ref{fig:correlations-1d-in-state-psi-beta}
for $\beta \in \{0.5, 2, 4, 8\}$.
We observe a long-range power-law decay that is
consistent with a power of $-2$ independent of $\beta$
and find that the correlator is negative at large distances.
This agrees with the term in the bosonization result
for the $XXZ$ model~\cite{Luther1975} that originates
from the current-current correlator.  The correlations
in $\psireal_{\beta}(\bm{s})$ do, however, not
have the staggered contribution
observed for the $XXZ$ model. We interpret this
as a smoothing effect due to the transition
from discrete to continuous spins.
A similar behavior was found in Ref.~\onlinecite{Herwerth2017}
in the context of approximating correlations for
spin-$\frac{1}{2}$ lattice states.

\subsection{Entanglement entropies}
\begin{figure*}[htb]
\centering
\includegraphics{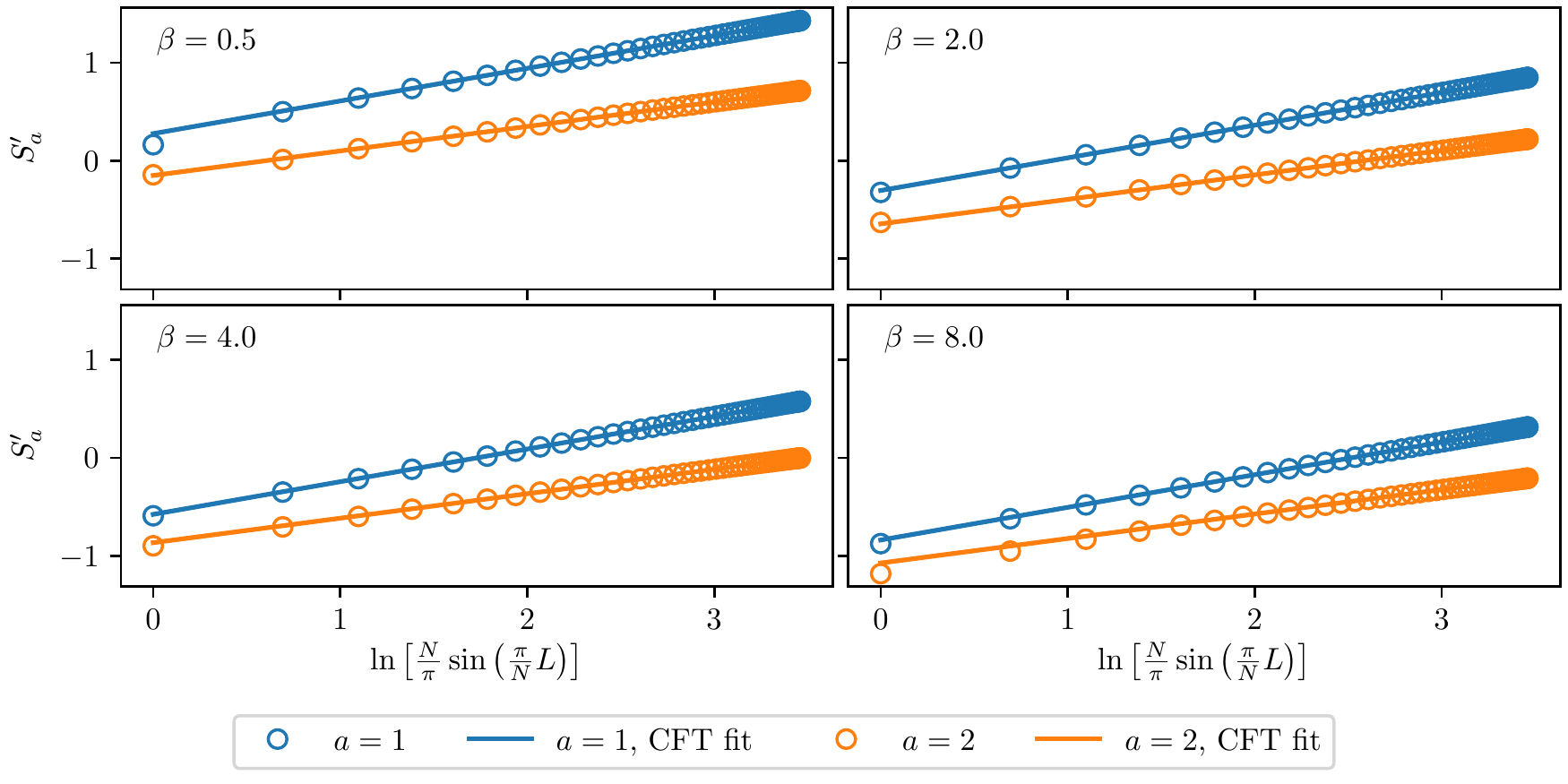}
\caption{(Color online) Entanglement entropies in $\psireal_{\beta}(\bm{s})$ in 1D
with periodic boundary conditions and $N=100$ sites.
The long-range behavior is consistent with
the CFT expectation of Eq.~\eqref{eq:label-CFT-entanglement-entropy}.
The quantities $S_a'$ can become negative since they differ from the R\'{e}nyi entropies $S_a$ by a divergent term, cf. Eq.~\eqref{eq:Sa-and-finite-Sap}. For the fit to the CFT formula,
we chose $c=1$ and used the $10$ data points with the largest value of $L$.
}
\label{fig:entanglement-entropy-1d-in-state-psi-beta}
\end{figure*}
For a partition of the system into two connected regions
of length $L$ and $N-L$, respectively,
the CFT entanglement entropy is given
by~\cite{Holzhey1994, Calabrese2004, Calabrese2009}
\begin{align}
\label{eq:label-CFT-entanglement-entropy}
S^{\mathrm{CFT}}_a(L) &= \frac{c}{6} \left(1 + \frac{1}{a}\right) \ln\left[ \frac{N}{\pi} \sin\left(\frac{\pi}{N} L\right)\right] + c_a',
\end{align}
where $c$ is the central charge,
$a$ the order of the R\'{e}nyi entropy, and $c_a'$ a non-universal constant.
As shown in Fig.~\ref{fig:entanglement-entropy-1d-in-state-psi-beta},
we find good agreement between the entropy of $\psireal_{\beta}(\bm{s})$
and $S^{\mathrm{CFT}}_a(L)$ for larger values of $L$.

For a system that has a low-energy description in terms of a Luttinger liquid,
one expects~\cite{Calabrese2010} subleading, oscillatory corrections to the CFT behavior of
Eq.~\eqref{eq:label-CFT-entanglement-entropy}. For the $XXZ$ model, for example, these
oscillations were found~\cite{Calabrese2010} in R\'{e}nyi entropies $S_a$ for $a \neq 1$.
From Fig.~\ref{fig:entanglement-entropy-1d-in-state-psi-beta},
we conclude that such oscillations around the CFT expectation
are absent for the state $\psireal_{\beta}(\bm{s})$.
This is in agreement with
our findings about the correlations, and we interpret it as a result of
the transition from discrete to continuous spins, which has a smoothing
effect and thus removes the oscillatory components.
We analyzed the deviation of the entanglement entropy in
$\psireal_{\beta}(\bm{s})$ from $S^{\mathrm{CFT}}_a(L)$.
For large distances $L$, we find a correction to the
CFT proportionality constant $\frac{c}{6} (1+\frac{1}{a})$.
This deviation becomes smaller for larger systems and
can thus be considered to be a finite size effect.

\subsection{Entanglement spectrum}
\begin{figure*}[htb]
\centering
\includegraphics{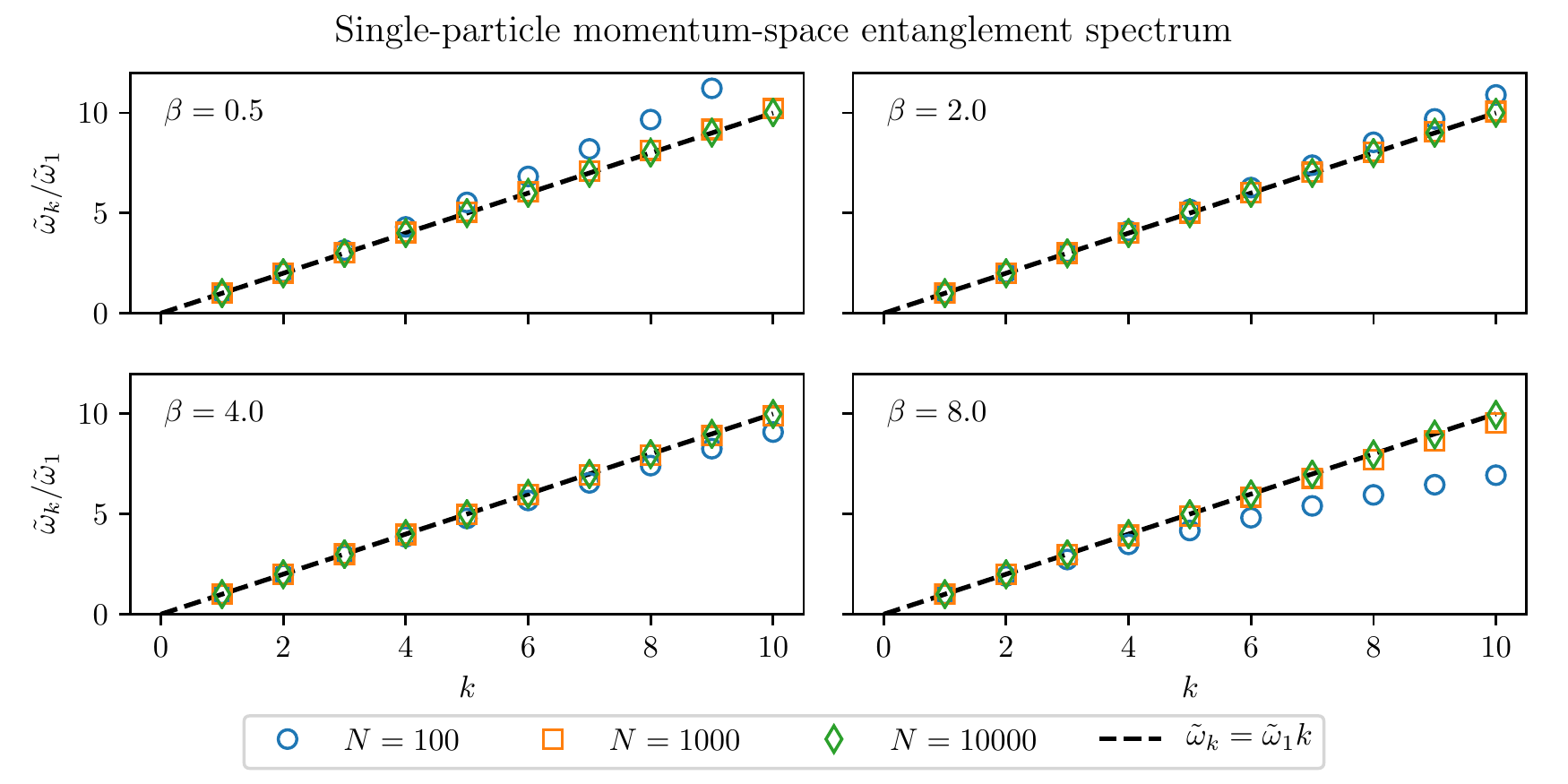}
\caption{(Color online) Single-particle entanglement spectrum in 1D with periodic
boundary conditions for a partition in momentum space. The low-lying part
of the spectrum is shown for various system sizes and values of $\beta$.
}
\label{fig:entanglement-spectrum-1d-in-state-psi-beta}
\end{figure*}
To further substantiate the close connection between $\psireal_{\beta}(\bm{s})$
and the free-boson CFT, we computed the entanglement
spectrum for a partition in momentum space.  This choice makes it possible to trace
out the negative momenta, thus retaining only the chiral components~\cite{Thomale2010, Lundgren2014}.
The details of the computation can be found in Appendix~\ref{sec:momentum-space-entanglement-spectrum-in-1D-psi-beta}. In summary, we find an entanglement Hamiltonian
$\sum_{k=1}^{\lfloor \frac{N - 1}{2} \rfloor} \tilde{\omega}_k b^\dagger_k b_k$, where $b_k$ and $b_k^\dagger$ are bosonic annihilation and creation operators, $k$ is the momentum, and
$\tilde{\omega}_k$ are single-particle entanglement energies.
We plot the low-lying part of the spectrum
in Fig.~\ref{fig:entanglement-spectrum-1d-in-state-psi-beta}.
For large systems, we observe a linear behavior $\tilde{\omega}_k = k \tilde{\omega}_1$,
where $\tilde{\omega}_1$ is the energy at momentum $k=1$.
The entanglement spectrum
is thus consistent with a chiral, massless, free boson.

\subsection{Parent Hamiltonian}
\begin{figure*}[htb]
\centering
\includegraphics{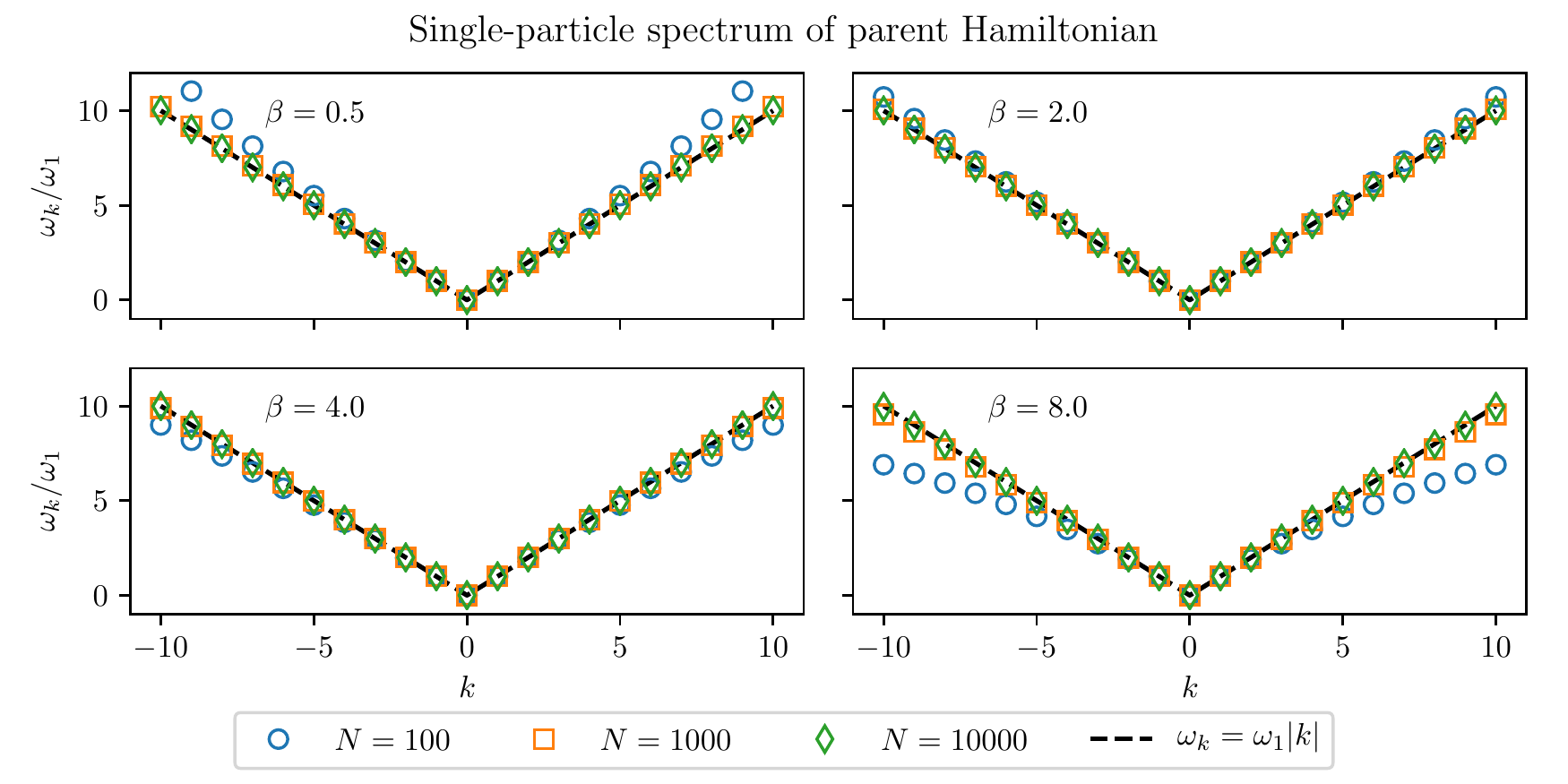}
\caption{(Color online) Low-lying energies of the single-particle spectrum for the parent Hamiltonian of $\psireal_{\beta}(\bm{s})$
in 1D with periodic boundary conditions.}
\label{fig:parent-Hamiltonian-psi-beta-tilde-spectrum-1D}
\end{figure*}
We now show that $\psireal_{\beta}(\bm{s})$ in 1D
has a parent Hamiltonian whose low-lying energies
are in agreement with the underlying CFT.
The precise form of this Hamiltonian is derived
in Appendix~\ref{appendix:parent-Hamiltonian-psireal-cont-s},
where we also show that it can be brought into the
diagonal form $\sum_{k=1}^N \omega_k b^\dagger_k b_k$
in a suitable basis of annihilation and creation
operators $b_k$ and $b_k^\dagger$. Due to translational
invariance, $k$ in the single-particle energies $\omega_k$
has the meaning of a momentum
variable in 1D.

The low-lying part of the single-particle spectrum is
shown in Fig.~\ref{fig:parent-Hamiltonian-psi-beta-tilde-spectrum-1D}.
The observed
linear behavior $\omega_k = \omega_1 |k|$ is consistent with CFT and our findings
about the momentum-space entanglement spectra. In the latter case, however,
the spectrum has only chiral components since the negative momenta were
traced out. 

\section{Properties of states in 2D}
\label{sec:properties-of-states-in-2D}
In this section, we consider a cylinder of size $N_x \times N_y$
as defined in Sec.~\ref{sec:cont-spin-states-cylinder}.
Through a determination of the topological entanglement entropy
and entanglement spectra, we provide evidence
that $\psireal_{\beta}(\bm{s})$ exhibits edge modes.

\subsection{Correlations}
Since the spin-spin correlations do not depend on the phase
of the wave function, the correlators
in $\psireal_{\beta}(\bm{s})$ agree with those of the
approximation made in Ref.~\onlinecite{Herwerth2017}, cf.
Appendix~\ref{appendix:relation-to-previous-approximation}.
In agreement with Ref.~\onlinecite{Herwerth2017},
we find an exponential decay of spin-spin correlations
in the bulk of a 2D system.
The long-range edge correlations decay with a power of
$-2$ independent of $\beta$, which agrees
with the decay of a current-current correlator
of the underlying CFT.

\subsection{Absence of intrinsic topological order}
The topological order of a state can be characterized
by the topological entanglement
entropy~\cite{Kitaev2006, Levin2006} $\gamma_{\mathrm{top}}$,
which occurs in the dependence of the entanglement entropy
$S_a(A)$ on the region $A$:
\begin{align}
\label{eq:area-law-of-entanglement-in-2D}
S_a(A) &= -\gamma_{\mathrm{top}} + b\; \partial A + \dots,
\end{align}
where $\partial A$ is the perimeter of $A$,
$b$ is a non-universal constant, and the dots stand for terms
that vanish for $\partial A \to \infty$.
A nonzero value of $\gamma_{\mathrm{top}}$ indicates
that a state exhibits intrinsic topological order.

\begin{figure*}[htb]
	\centering
	\includegraphics{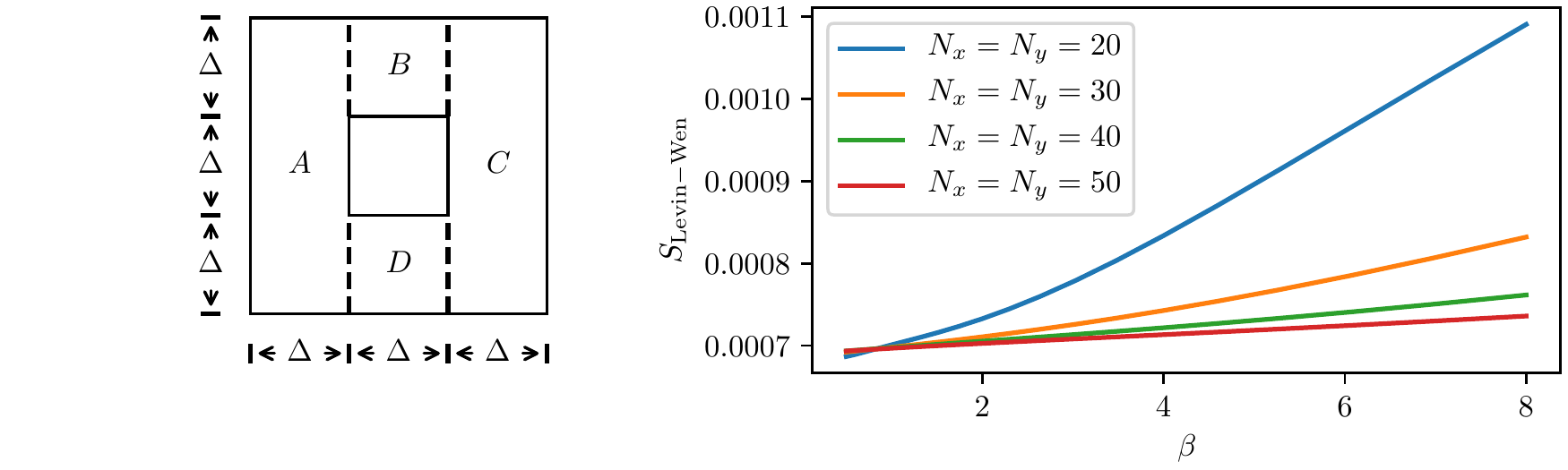}
	\caption{(Color online) Left panel: Definition of regions for the computation
		of the topological entanglement entropy according to Ref.~\onlinecite{Levin2006}. The regions $A$ and $C$
		are of size $\Delta \times 3 \Delta$ and the regions $B$
		and $D$ of size $\Delta \times \Delta$. We place the region $ABCD$
		into the center of a cylinder of size $N_x \times N_y$.
		Right panel: Linear combination $S_{\mathrm{Levin-Wen}}$
		of Eq.~\eqref{eq:definition-S-Levin-Wen} for different
		system sizes and values of $\beta$. The size of $\Delta$
		was chosen as $\Delta = N_x / 5$.
		}
	\label{fig:Levin-Wen-entropy}
\end{figure*}
The topological entanglement entropy can be computed as
a linear combination of entropies
for geometries that are chosen so that the
terms dependent on $\partial A$ in
Eq.~\eqref{eq:area-law-of-entanglement-in-2D}
drop out~\cite{Levin2006, Kitaev2006}.
Here, we consider the construction
of Levin and Wen~\cite{Levin2006} with regions as defined in the left panel of Fig.~\ref{fig:Levin-Wen-entropy}.
For geometries that are large compared to the correlation length, $\gamma_{\mathrm{top}}$ is equal to
\begin{align}
\label{eq:definition-S-Levin-Wen}
S_{\mathrm{Levin-Wen}} &= \frac{1}{2}
\big[
\left(S_1(ABC) - S_1(AC)\right)\\
&\phantom{ \frac{1}{2}\big[}
- \left(S_1(ABCD) - S_1(ADC)\right)
\big],\notag
\end{align}
where the R\'{e}nyi index $a=1$ was chosen.

We plot  $S_{\mathrm{Levin-Wen}}$ for $\psireal_{\beta}(\bm{s})$
in the right panel of Fig.~\ref{fig:Levin-Wen-entropy}.
For all considered system sizes and values of $\beta$, we observe
that $S_{\mathrm{Levin-Wen}}$ is below $0.002$. Furthermore,
$S_{\mathrm{Levin-Wen}}$ tends to decrease for larger systems.
This indicates that the state has a vanishing
topological entanglement entropy, $\gamma_{\mathrm{top}} = 0$,
and thus no intrinsic topological order.
A similar observation of a vanishing topological
entanglement entropy was made for
BCS states with a $p_x + i p_y$ symmetry
in Refs.~\onlinecite{Montes2017b, BrayAli2009}.

\subsection{Entanglement spectrum and edge states}
\label{sec:psi-cont-2d-entanglement-spec}
\begin{figure}[htb]
	\centering
	\includegraphics{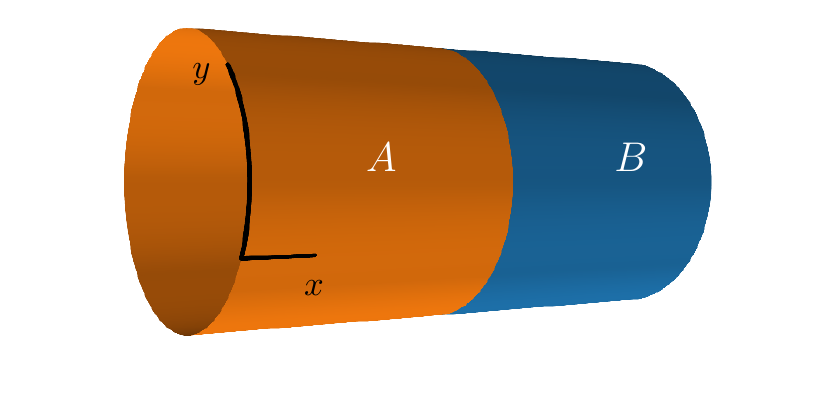}
	\caption{(Color online) Cut of the cylinder into two pieces
	$A$ and $B$ for the computation of the entanglement spectrum.}
	\label{fig:cylinder-cut}
\end{figure}
In the previous subsection, we provided evidence that $\psireal_{\beta}(\bm{s})$ does not have intrinsic topological
order. We now study the entanglement spectrum and
show that it contains indications of edge states.

\begin{figure*}[htb]
	\centering
	\includegraphics{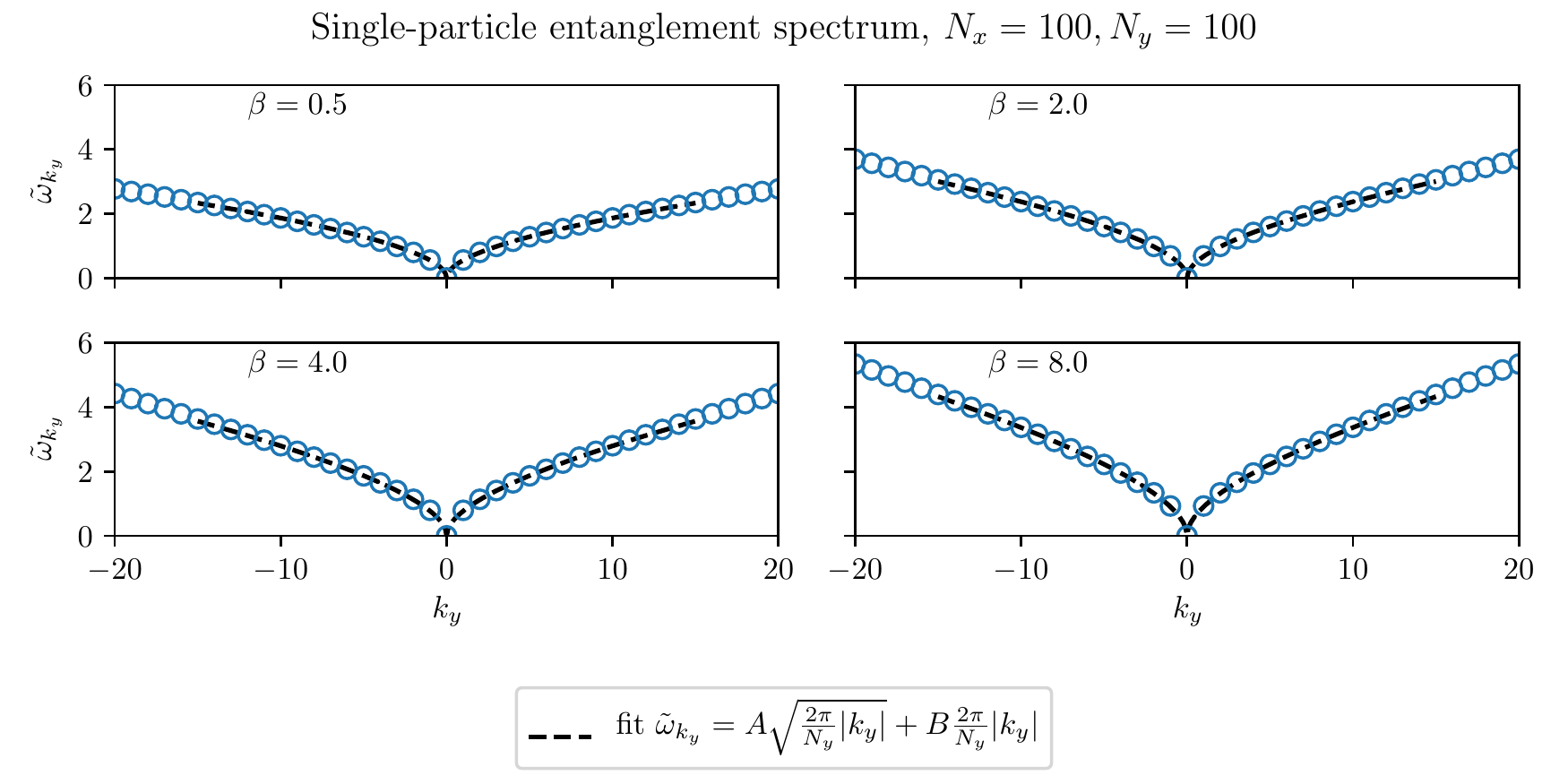}
	\caption{(Color online) Low-lying part of the single-particle entanglement spectrum
		in 2D for a cut of the cylinder into two
		pieces as shown in Fig.~\ref{fig:cylinder-cut}.
	}
	\label{fig:2D-entanglement-spectrum-psi-cont}
\end{figure*}
We consider a partition of the cylinder
into two pieces of equal size, where we choose the
cut perpendicular to the open direction as illustrated
in Fig.~\ref{fig:cylinder-cut}.
The low-lying part of the single-particle entanglement
spectrum
is shown in Fig.~\ref{fig:2D-entanglement-spectrum-psi-cont}.
Since the cut preserves translational symmetry,
the spectrum can be ordered according to the momentum $k_y$
in the periodic direction.
The dependence of the low-lying
single-particle energies $\tilde{\omega}_{k_y}$ is consistent
with
\begin{align}
\label{eq:tilde-omega-sqrt-ky}
\tilde{\omega}_{k_y} &= A \sqrt{\frac{2 \pi}{N_y} |k_y|} +  B \frac{2 \pi}{N_y} |k_y|,
\end{align}
where $A$ and $B$ are fit constants. For large systems,
we find that $A$ and $B$ are independent of the system
size within variations that are due to the chosen fit range.
Thus, $\omega_{k_y} \propto \sqrt{|k_y|}$
for the smallest momenta.
A similar dispersion relation was recently found
in the entanglement spectra of coupled Luttinger
liquids~\cite{Lundgren2013}.

Next, we investigate whether the low-lying excited states
in the entanglement spectrum are localized at the boundary
created by the cut and thus represent edge excitations.
To this end, we compute the basis change that makes
the entanglement Hamiltonian diagonal.
To exploit translational symmetry, we use
Fourier transformed annihilation and creation
operators $\tilde{a}_{i_x k_y}$ and $\tilde{a}_{i_x k_y}^\dagger$.
As shown in Appendix~\ref{sec:appendix-entanglement-spec-cylinder-fourier-transform},
the entanglement Hamiltonian is diagonal in annihilation and creation
operators $b_{i_x l_y \sigma}$ and $b_{i_x l_y \sigma}^\dagger$,
where $l_y$
is an index of non-negative momenta and $\sigma$ a sign index
($\sigma \in \{+,-\}$ for $l_y \notin \{0, \frac{N_y}{2}\}$
and $\sigma = +$ for $l_y \in \{0, \frac{N_y}{2}\}$).
For $l_y \notin \{0, \frac{N_y}{2}\}$, the transformation
to the diagonal basis assumes the form
\begin{align}
    \label{eq:basis-transformation-a-to-b-for-cylinder-cut}
\left(\begin{matrix}
\bm{b}_{l_y, \sigma=+}\\
\bm{b}^\dagger_{l_y, \sigma=+}
\end{matrix}\right) &= R_{l_y} \frac{1}{\sqrt{2}}\left(\begin{matrix}
\tilde{\bm{a}}_{l_y} + \tilde{\bm{a}}_{-l_y}\\
\tilde{\bm{a}}_{l_y}^\dagger + \tilde{\bm{a}}_{-l_y}^\dagger\\
\end{matrix}\right), \\
\left(\begin{matrix}
\bm{b}_{l_y, \sigma=-}\\
\bm{b}^\dagger_{l_y, \sigma=-}
\end{matrix}\right) &= R_{l_y} \frac{1}{\sqrt{2}}\left(\begin{matrix}
-i \tilde{\bm{a}}_{l_y} + i \tilde{\bm{a}}_{-l_y}\\
i \tilde{\bm{a}}_{l_y}^\dagger -i \tilde{\bm{a}}_{-l_y}^\dagger\\
\end{matrix}\right),
\end{align}
where $\bm{b}_{l_y \sigma} = (b_{i_x=1, l_y, \sigma}, \dots, b_{i_x=\frac{N_x}{2}, l_y, \sigma})^t$
and analogously for the other annihilation and creation operators.
The $N_x \times N_x$ matrix $R_{l_y}$ is a symplectic basis transformation
in the complex representation,
\begin{align}
R_{l_y} &= \left(\begin{matrix}
R^{(1)}_{l_y} & R^{(2)}_{l_y}\\
\left(R^{(2)}_{l_y}\right)^{*} & \left(R^{(1)}_{l_y}\right)^{*}
\end{matrix}\right).
\end{align}
We order the spectrum so that $(R^{(r)}_{l_y})_{i_x j_x}$
with $i_x = 1$ and $r \in \{1, 2\}$
corresponds to the lowest energy state in the sector of momentum $l_y$.
According to Eq.~\eqref{eq:basis-transformation-a-to-b-for-cylinder-cut},
$\bm{b}_{l_y \sigma}$ are linear
combinations of modes with momenta $l_y$ and $-l_y$.
The two choices $\sigma \in \{+, -\}$ have the same
energy and correspond to the degeneracies
in Fig.~\ref{fig:2D-entanglement-spectrum-psi-cont}.

\begin{figure}[htb]
	\centering
	\includegraphics{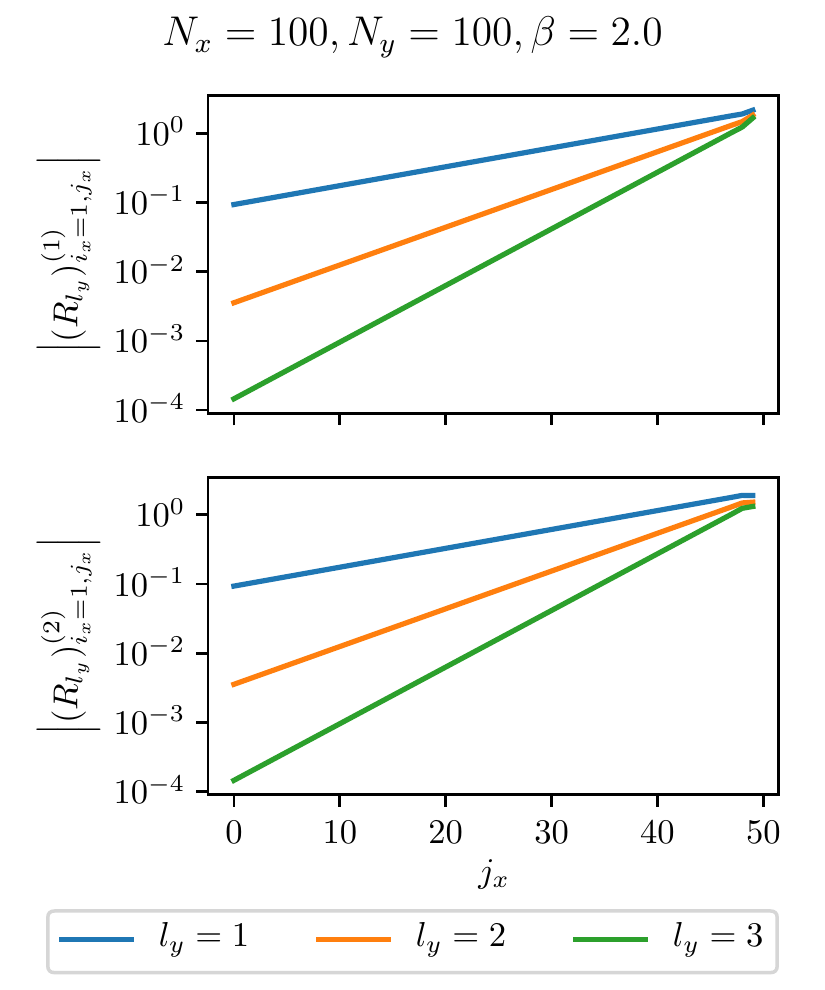}
	\caption{(Color online) Amplitude of the basis transformation for the first excited
	states in the entanglement spectrum of Fig.~\ref{fig:2D-entanglement-spectrum-psi-cont}. The shown data is for $\beta = 2.0$. (The corresponding amplitudes for $\beta \in \{0.5, 4, 8\}$
	have the same qualitative behavior.)
	The lowest energy
	in the sector of momentum $l_y$ corresponds to $i_x=1$, cf. the explanation
	below Eq.~\eqref{eq:basis-transformation-a-to-b-for-cylinder-cut}.
	The value $j_x = 50$ is the position of the cut.
	}
	\label{fig:2D-entanglement-spectrum-psi-cont-basis-change}
\end{figure}
Fig.~\ref{fig:2D-entanglement-spectrum-psi-cont-basis-change} shows $(R^{(r)}_{l_y})_{i_x j_x}$
for $i_x = 1$ and $l_y \in \{1, 2, 3\}$.
We observe that $(R^{(r)}_{l_y})_{i_x j_x}$
falls off exponentially in $j_x$ for large distances
to the position of the cut ($j_x=50$). Thus, the corresponding
states are exponentially localized at the edge created
by the cut.
This observation provides evidence that $\psireal_{\beta}(\bm{s})$
indeed supports gapless edge states.

We also did analogous
computations for the local parent Hamiltonian of Appendix~\ref{appendix:parent-Hamiltonian-psireal-cont-s}
to test whether its low-lying
excited states are localized at the physical edges.
Compared to the entanglement spectrum, we observed the
following differences. First, the low-lying single-particle
spectrum of the parent Hamiltonian does not consist of a single branch
as the entanglement spectrum of Fig.~\ref{fig:2D-entanglement-spectrum-psi-cont}.
Second, we find eigenstates of the Hamiltonian with low energies
that are not localized at the edge. This raises the
question whether there is another parent Hamiltonian with
the same low-energy behavior as observed in the entanglement
spectrum.

\section{Conclusion}
\label{sec:conclusion}
This paper considers continuous-spin
wave functions $\psireal_{\beta}(\bm{s})$ on lattices
that are constructed as correlators of the massless, free boson CFT.
In contrast to the case of
discrete spins or continuous positional degrees of freedom,
the wave functions $\psireal_{\beta}(\bm{s})$ are Gaussian
and their properties can be computed
efficiently using the formalism of bosonic Gaussian states.

Through an analysis of entanglement entropies
and spectra, we found that
$\psireal_{\beta}(\bm{s})$ is closely related
to the underlying CFT in 1D.
More precisely, we observed a good agreement between the
entanglement entropy of $\psireal_{\beta}(\bm{s})$  and the CFT expectation. In contrast to some lattice systems like
the $XXZ$ model~\cite{Calabrese2010}, we do not find
subleading oscillatory corrections to the CFT behavior.
At small energies, we recovered the underlying CFT of a free,
massless boson, in the momentum space entanglement spectrum
and also in the spectrum of a parent Hamiltonian.

In 2D, we probed possible topological properties
of $\psireal_{\beta}(\bm{s})$ through an analysis
of entanglement entropies and spectra.  Although our results are consistent with
a vanishing topological entanglement entropy, we
found evidence for edge states in the entanglement
spectrum. The absence of intrinsic topological
order is distinct from the chiral case with discrete spins.

The wave function $\psireal_{\beta}(\bm{s})$ is real
since it is constructed from the full bosonic field
$\varphi(z, \bar{z})$. As a consequence, the entanglement
Hamiltonian in 2D has eigenstates that are linear combinations
of left- and right-moving modes.
Together with our observation
of states localized at the edge in the low-lying entanglement spectrum,
this is an indication that  $\psireal_{\beta}(\bm{s})$ could
describe a state that is similar to a quantum spin Hall effect.

We found a local parent Hamiltonian whose low-lying energy
levels in 1D are consistent with the corresponding entanglement
spectrum. In 2D, however, this parent Hamiltonian
has low-lying excited states that are not localized at the edge,
which is in contrast to the entanglement spectrum.
It would be interesting to investigate if there is another local parent
Hamiltonian with the same low-energy properties as observed
in the entanglement spectrum.

In 1D, the real wave function $\psireal_{\beta}(\bm{s})$ is equivalent
to the analogously defined chiral wave function,
which is constructed from the chiral part of the free-boson
field.
For general lattice configurations, however, $\psireal_{\beta}(\bm{s})$
differs from its chiral counterpart. In contrast to the
real case, we found that the chiral state depends
on the ordering of the lattice
positions. It could, therefore, be that another framework than that
of bosonic Gaussian states is necessary to consistently
treat the chiral state, and it would be interesting
to investigate this question in a future study.

\begin{acknowledgments}
	We are grateful to Tao Shi, Norbert Schuch, and Ivan Glasser for discussions.
	This work was supported by
	the Spanish government program FIS2015-69167-C2-1-P,
	the Comunidad de Madrid grant QUITEMAD+ S2013/ICE-2801,
	the grant SEV-2016-0597 of the Centro de Excelencia Severo Ochoa Programme,
	the DFG within the Cluster of Excellence NIM,
	and the ERC grant QUENOCOBA, ERC-2016-ADG (Grant No. 742102).
\end{acknowledgments}

\appendix

\section{The formalism of Gaussian states}
\label{sec:appendix-general-Gaussian-states}
In this section, we review the formalism of bosonic Gaussian states
as needed throughout this paper. In particular, we
describe how to compute bipartite entanglement properties of a generic,
pure Gaussian state.

We consider Gaussian wave functions
\begin{align}
\label{eq:general-pure-state-gaussian-wave-function}
\psi(\bm{s}) &= e^{-\frac{1}{2} \bm{s}^t(X + i Y) \bm{s}},
\end{align}
where $\bm{s} = (s_1, \dots, s_N)^t \in \mathbb{R}^N$ and $X$
and $Y$ are real $N \times N$ matrices. Normalizability of the
wave function $\psi(\bm{s})$ requires that $X > 0$.
We think of $\bm{s}$ as the vector of $N$ continuous, effective spin
variables.

The wave function $\psi(\bm{s})$ has the same form as that of $N$ continuous, positional degrees on freedom on the real line.
Therefore, we introduce the operators
$\bm{Q} = (Q_1, \dots, Q_N)^t$ and $\bm{P} = (P_1, \dots, P_N)^t$
through
\begin{align}
(Q_m \psi)(\bm{s}) &= s_m \psi(\bm{s}),&\text{and }
(P_m \psi)(\bm{s}) &= -i \frac{\partial \psi}{\partial s_m}(\bm{s}).
\end{align}

The canonical commutation relations $\left[Q_m, P_n\right] = i \delta_{m n}$ can be written as
\begin{align}
\label{eq:commutation-relations-of-F}
\left[R_m, R_n\right] &= i \Omega_{m n},
\end{align}
where $\bm{R} = (Q_1, \dots, Q_N, P_1, \dots, P_N)^t$, and $\Omega$
is the $2 N \times 2 N$ matrix
\begin{align}
\Omega &= \left(\begin{matrix}
0 & \mathbb{I}\\
-\mathbb{I} & 0
\end{matrix}\right)
\end{align}
with $\mathbb{I}$ being the $N \times N$ identity matrix.
The bosonic creation and annihilation operators are defined as
\begin{align}
\left(\begin{matrix}
\bm{a}\\
\bm{a}^\dagger
\end{matrix}\right) &= \mathcal{U} \left(\begin{matrix}
\bm{Q}\\
\bm{P}
\end{matrix}\right),
\intertext{where}
\label{eq:psi-cont-u-for-transition-to-a}
\mathcal{U} &= \frac{1}{\sqrt{2}} \left(\begin{matrix}
\mathbb{I} & i \mathbb{I}\\
\mathbb{I} & -i \mathbb{I}
\end{matrix}\right),
\end{align}
$\bm{a} = \left(a_1, \dots, a_N\right)^t$, and
$\bm{a}^\dagger = \left(a_1^\dagger, \dots, a_N^\dagger\right)^t$.
These operators satisfy the canonical commutation relations
\begin{align}
\left[\left(\begin{matrix}
\bm{a}\\
\bm{a}^\dagger
\end{matrix}\right)_m, \left(\begin{matrix}
\bm{a}\\
\bm{a}^\dagger
\end{matrix}\right)_n\right] &= \Omega_{m n}.
\end{align}

Symplectic matrices $S$ are defined as $2 N \times 2 N$ real matrices
satisfying $S \Omega S^t = \Omega$. They preserve the commutation relations
of Eq.~\eqref{eq:commutation-relations-of-F}: Given a symplectic
matrix $S$, the vector $\bm{R}' = S \bm{R}$ satisfies
$\left[R'_m, R_n'\right] = \left[R_m, R_n\right] = i \Omega_{m n}$.

A Gaussian state $\psi(\bm{s})$ is completely characterized by its covariance matrix
\begin{align}
\label{eq:def-covariance-matrix}
\gamma_{m n} &= \frac{\langle \psi | \left\{R_m, R_n\right\} |\psi \rangle}{\langle \psi | \psi\rangle},
\end{align}
where $\left\{\bullet, \bullet\right\}$ is the anticommutator.

Using the wave function of Eq.~\eqref{eq:general-pure-state-gaussian-wave-function},
one finds the covariance matrix~\cite{Wolf2003, DeGosson2006}
\begin{align}
\label{eq:gamma-cov-pure-state}
\gamma &= \left(\begin{matrix}
X^{-1} & -X^{-1} Y\\
-Y X^{-1} & X + Y X^{-1} Y
\end{matrix}\right).
\end{align}

In this paper, we study entanglement properties under a partition of the
system into two parts. Given a bipartition $A = \{i_1, \dots, i_L\}$,
$B = \{j_1, \dots, j_{N-L}\}$ with $A \cup B = \{1, \dots, N\}$,
these are encoded in the reduced density matrix $\rho_A$ obtained from
the pure-state density matrix
\begin{align}
\frac{|\psi \rangle \langle \psi |}{\langle \psi| \psi\rangle}
\end{align}
by tracing out the degrees of freedom of subsystem $B$:
\begin{align}
\rho_A &= \mathrm{tr}_B \frac{|\psi \rangle \langle \psi |}{\langle \psi | \psi\rangle}.
\end{align}
In terms of the covariance
matrix, this operation assumes a particularly simple form. Namely, the
covariance matrix $\gamma_A$ in the state $\rho_A$ obtained by tracing out the degrees of freedom
in $B$ is given by removing the rows and columns corresponding to $B$
from $\gamma$.~\cite{NavarreteBenlloch2015}

The R\'{e}nyi entanglement entropy of order $a$ is defined as
\begin{align}
S_a(A) &= \frac{1}{1 - a} \ln \mathrm{tr}\left[ \left(\rho_A\right)^a\right],
\end{align}
where the limit $a \to 1$ corresponds to the von Neumann entropy.
In terms of the covariance matrix $\gamma_A$ of the reduced state,
the entropies $S_a$ can be computed as~\cite{Adesso2004}
\begin{align}
\label{eq:entanglement-entropies-in-terms-of-symplectic-ev}
S_a(A) &= \sum_{j=1}^{L} g_a(\nu_j),
\intertext{where}
\label{eq:definition-of-ga}
g_a(y) &= \frac{1}{a - 1} \ln\left[\left(\frac{y+1}{2}\right)^a - \left(\frac{y-1}{2}\right)^a\right].
\end{align}
Here, $\nu_k$ are the symplectic eigenvalues of the matrix $\gamma_A$,
which are the positive eigenvalues of $i \gamma_A \Omega$.
The von Neumann entropy is given by the limit $a \to 1$:
\begin{align}
 \lim_{a \to 1}g_a(y) = \frac{y+1}{2} \ln\left(\frac{y+1}{2}\right) -\frac{y-1}{2} \ln\left(\frac{y-1}{2}\right).
\end{align}

The entanglement Hamiltonian $H_A$ is defined as the Hamiltonian whose
thermal state is given by the reduced density matrix, $\rho_A = e^{-H_A}$.
The factorization of a covariance matrix in terms of
a symplectic basis transformation and a diagonal matrix consisting of symplectic eigenvalues corresponds to a decomposition into a product
state of thermal oscillators~\cite{Peschel2009, Weedbrook2012}.
Thus, the symplectic
spectrum of the reduced state's covariance matrix is directly related
to single-particle energies of an entanglement Hamiltonian. Each symplectic
eigenvalue $\nu_j$ corresponds to an energy
\begin{align}
\label{eq:relation-between-symplectic-ev-and-entanglement-spectrum}
\tilde{\omega}_j &= \ln \frac{\nu_j + 1}{\nu_j - 1}.
\end{align}

\section{Details on entanglement properties}
\label{sec:appendix-entanglement-properties-of-Gaussian-states-from-CFT}
This section explains how we compute entanglement properties
for the Gaussian wave function $\psireal_{\beta}(\bm{s})$
defined in Sec.~\ref{sec:Gaussian-states-definitions} of the main text.
With respect to the case of generic Gaussian states discussed in
Appendix~\ref{sec:appendix-general-Gaussian-states},
we now have to take into account the delta function in
$\psireal_{\beta}(\bm{s})$, which leads to divergences.
The regularization explained in
Sec.~\ref{sec:regularization-of-delta-function} leads
to the wave function $\psireal_{\beta, \epsilon}(\bm{s})$
with covariance matrix [cf. Eq.~\eqref{eq:gamma-cov-pure-state}]
\begin{align}
\label{eq:cov-matrix-in-regularized-wave-function}
\gamma_{\beta, \epsilon} &= \frac{1}{2 \epsilon} \left(\begin{matrix}
0 & 0\\
0 & \bm{e} \bm{e}^t
\end{matrix}\right) + \gamma'_{\beta, \epsilon},
\intertext{where}
\gamma'_{\beta, \epsilon} &= \left(\begin{matrix}
X_{\beta, \epsilon}^{-1} & 0\\
0 & X_{\beta}
\end{matrix}\right).
\end{align}
Using~\cite{Bartlett1951}
\begin{align}
\label{eq-inverse-of-Xinv-beta-epsilon}
X^{-1}_{\beta, \epsilon} &= X^{-1}_{\beta} - \frac{1}{2 \epsilon + \bm{e}^t X^{-1}_{\beta} \bm{e}} X^{-1}_{\beta} \bm{e} \bm{e}^t X^{-1}_{\beta},
\end{align}
we find that $\gamma_{\beta, \epsilon}'$ is finite in the limit $\epsilon \to 0$.
In particular, the $QQ$, $QP$, and $PQ$ blocks of the covariance matrix $\gamma_{\beta, \epsilon}$ are finite,
while the $PP$ block has a divergent term.

\subsection{Symplectic eigenvalues of the reduced state's covariance matrix}
\label{sec:appendix-symplectic-eigenvalues-of-reduced-state}
Let us write the covariance matrix of Eq.~\eqref{eq:cov-matrix-in-regularized-wave-function} as
\begin{align}
\gamma_{\beta, \epsilon} &= \frac{1}{2 \epsilon} v v^t + \gamma'_{\beta, \epsilon},
&\text{where }
v &= \left(\begin{matrix}
0\\
\bm{e}
\end{matrix}\right).
\end{align}

In the following, we consider a bipartition into disjoint subsystems $A = \{i_1, \dots, i_L\}$
and $B = \{j_1, \dots, j_{N-L}\}$, where $A \cup B = \{1, \dots, N\}$ and $L \in \{1, \dots, N-1\}$.
The covariance matrix after tracing out the subsystem $B$
is given by $\mathcal{D}_A \gamma_{\beta, \epsilon} \mathcal{D}_A^t$, where
\begin{align}
\mathcal{D}_A &= \left(\begin{matrix}
D_A & 0\\
0 & D_A
\end{matrix}\right)
\intertext{with the $L \times N$ matrix}
D_A &= \left(\begin{matrix}
\text{---}  \bm{e}_{i_1}^t  \text{---}\\
  \vdots  \\
\text{---}  \bm{e}_{i_L}^t  \text{---}
\end{matrix}\right)
\end{align}
and $\bm{e}_i$ being the $i$th unit vector.
The matrix $\mathcal{D}_A$ removes the rows and columns corresponding to $B$ from $\gamma_{\beta, \epsilon}$.

The entanglement entropies and spectra follow directly from
the symplectic eigenvalues of $\mathcal{D}_A \gamma_{\beta, \epsilon} \mathcal{D}_A^t$,
which are the positive eigenvalues of $i \Omega \mathcal{D}_A \gamma_{\beta, \epsilon} \mathcal{D}_A^t$.
However, the covariance matrix $\gamma_{\beta, \epsilon}$
is divergent in the limit $\epsilon \to 0$, which leads to an infinity in the
symplectic eigenvalues and thus
in the entropies. To handle this divergence, we compute the inverse $\left[\mathcal{D}_A \gamma_{\beta, \epsilon} \mathcal{D}_A^t\right]^{-1}$ since it is finite for $\epsilon \to 0$:
\begin{widetext}
\begin{align}
\left[\mathcal{D}_A \gamma_{\beta, \epsilon} \mathcal{D}_A^t\right]^{-1} &= \left[\mathcal{D}_A \gamma_{\beta, \epsilon}' \mathcal{D}^t_A\right]^{-1} - \frac{\left[\mathcal{D}_A \gamma_{\beta, \epsilon}' \mathcal{D}_A^t\right]^{-1} \mathcal{D}_A v v^t \mathcal{D}_A^t \left[\mathcal{D}_A \gamma_{\beta, \epsilon}' \mathcal{D}_A^t\right]^{-1}}{2 \epsilon + v^t \mathcal{D}_A^t  \left[\mathcal{D}_A \gamma_{\beta, \epsilon}' \mathcal{D}^t\right]^{-1} \mathcal{D}_A v},
\end{align}
where we used the formula of Ref.~\onlinecite{Bartlett1951} to
compute the inverse of a matrix that is changed by a term of rank one.
With $\mu^{(1)}_{\beta, \epsilon} \leq \mu^{(2)}_{\beta, \epsilon}
\dots \leq \mu^{(L)}_{\beta, \epsilon}$ being the ordered positive
eigenvalues of $-i \left[\mathcal{D}_A \gamma_{\beta, \epsilon}
\mathcal{D}_A^t\right]^{-1} \Omega$, the symplectic spectrum of
$\mathcal{D}_A \gamma_{\beta, \epsilon} \mathcal{D}_A^t$ is then
given by
$\{\nu^{(j)}_{\beta, \epsilon} = 1/\mu^{(j)}_{\beta, \epsilon}\}_{1 \leq j \leq {L}}$.
For $\epsilon \to 0$, we have $\mu^{(1)}_{\beta, \epsilon} \to 0$ so
that $\nu^{(1)}_{\beta, \epsilon} \to \infty$.

Let us now compute how $\nu^{(1)}_{\beta, \epsilon}$ scales with $\epsilon$ for $\epsilon \to 0$.
This will be needed to subtract the divergence from the resulting entanglement
entropy.
With $\det \Omega = 1$, we have
\begin{align}
\ln \det \mathcal{D}_A \gamma_{\beta, \epsilon} \mathcal{D}_A^t
&=\ln \det \Omega \mathcal{D}_A \gamma_{\beta, \epsilon} \mathcal{D}_A^t = 
\sum_{j=1}^{L} \ln\left[\left(\nu^{(j)}_{\beta, \epsilon}\right)^2\right]
\end{align}
and therefore
\begin{align}
\ln \nu^{(1)}_{\beta, \epsilon} &= \frac{1}{2} \ln \det \mathcal{D}_A \gamma_{\beta, \epsilon} \mathcal{D}_A^t  -\sum_{j=2}^{L} \ln\left[ \nu^{(j)}_{\beta, \epsilon}\right].
\end{align}
The matrix determinant lemma allows to express the determinant
of $\mathcal{D}_A \gamma_{\beta, \epsilon} \mathcal{D}_A^t$ in
terms of $\mathcal{D}_A \gamma_{\beta, \epsilon}' \mathcal{D}_A^t$,
which differs from $\mathcal{D}_A \gamma_{\beta, \epsilon} \mathcal{D}_A^t$
only by a term of rank one.
Thus, we obtain
\begin{align}
\label{eq:divergent-symplectic-eigenvalue}
\ln \nu^{(1)}_{\beta, \epsilon}
&= \frac{1}{2} \ln \left[\left(1 + \frac{1}{2 \epsilon} v^t \mathcal{D}_A^t \left(\mathcal{D}_A \gamma_{\beta, \epsilon}' \mathcal{D}_A^t\right)^{-1} \mathcal{D}_A v\right) \det\left(\mathcal{D}_A \gamma_{\beta, \epsilon}' \mathcal{D}_A^t\right)\right]  -\sum_{j=2}^L \ln \nu^{(j)}_{\beta, \epsilon}\\
\label{eq:divergent-symplectic-eigenvalue-line2}
&= -\frac{1}{2} \ln \epsilon + \ln \tilde{\nu}_{\beta} + \mathcal{O}(\epsilon),
\intertext{where}
 \ln \tilde{\nu}_\beta  &= \frac{1}{2} \ln\left[\frac{\det\left(\mathcal{D}_A \gamma_{\beta, \epsilon=0}' \mathcal{D}_A^t\right)}{2} v^t \mathcal{D}_A^t \left(\mathcal{D}_A \gamma_{\beta, \epsilon=0}' \mathcal{D}_A^t\right)^{-1} \mathcal{D}_A v \right]  -\sum_{j=2}^{L} \ln \nu^{(j)}_{\beta, \epsilon = 0}.
\end{align}
\end{widetext}

\subsection{Entanglement entropies and spectra}
\label{sec:entanglement-entropies-and-spectra-psi-beta}
Having computed the symplectic
spectrum $\{\nu^{(1)}_{\beta, \epsilon}, \dots, \nu^{(L)}_{\beta, \epsilon}\}$ of the reduced state's covariance matrix,
we can now determine the entanglement properties in $\psireal_{\beta}(\bm{s})$.

The divergent symplectic eigenvalue assumes the form
$\ln \nu_{\beta, \epsilon}^{(1)} = -\frac{1}{2} \ln \epsilon + \ln \tilde{\nu}_{\beta} + \mathcal{O}(\epsilon)$
[cf. Eq.~\eqref{eq:divergent-symplectic-eigenvalue-line2}],
and thus we find
\begin{align}
g_a(\nu^{(1)}_{\beta, \epsilon}) &= -\frac{1}{2} \ln(\epsilon) + \frac{1}{a-1} \ln(a) - \ln(2)   + \ln \tilde{\nu}_\beta\\
&\quad + \mathcal{O}(\epsilon),\notag
\end{align}
where $g_a(y)$ was defined in Eq.~\eqref{eq:definition-of-ga}.
From Eq.~\eqref{eq:entanglement-entropies-in-terms-of-symplectic-ev},
the entanglement entropies then follow as
\begin{align}
S_a(A) &= -\frac{1}{2} \ln \epsilon + S_a'(A) + \mathcal{O}(\epsilon),
\intertext{where}
S_a'(A) &= \frac{1}{a-1} \ln(a) - \ln(2)   + \ln \tilde{\nu}_{\beta} + \sum_{j=2}^{L} g_a\left(\nu^{(j)}_{\beta, \epsilon=0}\right).
\end{align}
The entropy $S_a'(A)$ differs from $S_a(A)$ by the subtraction
of the divergent term $-\frac{1}{2} \ln \epsilon$.

Given the symplectic eigenvalues $\nu^{(j)}_{\beta, \epsilon}$, one can also compute the entanglement spectrum. According to Eq.~\eqref{eq:relation-between-symplectic-ev-and-entanglement-spectrum}, we find the single-particle entanglement energies
\begin{align}
\tilde{\omega}_j &= \begin{cases}
0 &\text{if } j = 1,\\
\ln \frac{\nu^{(j)}_{\beta, \epsilon=0} + 1}{\nu^{(j)}_{\beta, \epsilon=0}  - 1}&\text{if } j \neq 1
\end{cases}
\end{align}
in the limit $\epsilon \to 0$. Since $\nu^{(1)}_{\beta, \epsilon} \to \infty$ for $\epsilon \to 0$, the energy $\tilde{\omega}_1$ vanishes.
The entanglement Hamiltonian thus assumes the form
$\sum_{j=1}^L \tilde{\omega}_j b^\dagger_j b_j$, where
$b_j$ and $b^\dagger_j$ are annihilation and creation
operators in a suitable basis. The precise relation
between the original operators $(a_j, a_j^\dagger)$
and $(b_j, b_j^\dagger)$ can be determined
by computing Williamson's normal form~\cite{Williamson1936, DeGosson2006}
of the covariance matrix corresponding to the state's reduced
density matrix.

\subsection{Momentum-space entanglement spectrum in 1D with periodic boundary conditions}
\label{sec:momentum-space-entanglement-spectrum-in-1D-psi-beta}
We now consider the state $\psireal_{\beta}(\bm{s})$ in 1D with periodic
boundary conditions. For this choice, the matrix $\left(X_{\beta, \epsilon}\right)_{i, j}$ only depends on the difference $i - j$ modulo $N$,
and thus we write $\left(X_{\beta, \epsilon}\right)_{i, j} = \left(X_{\beta, \epsilon}\right)_{i - j}$.

We consider the discrete Fourier transform
\begin{align}
F_{k j} &= \frac{1}{\sqrt{N}} e^{-2 \pi i \frac{k j}{N}},
\end{align}
where the normalization was chosen so that $F$ is unitary.

Writing $F = F_x + i F_y$ with $F_x$ and $F_y$ real, we define the symplectic transformation
\begin{align}
\mathcal{F} &= \left(\begin{matrix}
F_x  & -F_y\\
F_y & F_x
\end{matrix}\right),
\end{align}
which corresponds to a unitary rotation of creation and annihilation operators
of the form
\begin{align}
\left(\begin{matrix}
\bm{a}\\
\bm{a}^\dagger
\end{matrix}\right) &\to \left(\begin{matrix}
F  & 0\\
0 & F^{*}
\end{matrix}\right) \left(\begin{matrix}
\bm{a}\\
\bm{a}^\dagger
\end{matrix}\right).
\end{align}
Therefore, $\mathcal{F}$ is the symplectic matrix that transforms to momentum space.

\begin{widetext}
The covariance matrix of Eq.~\eqref{eq:cov-matrix-in-regularized-wave-function} transformed to momentum space then becomes
\begin{align}
\mathcal{F} \gamma_{\beta, \epsilon} \mathcal{F}^t &= \left(\begin{matrix}
\gamma_{\beta, \epsilon}^{(1)} & 0 \\
0 & \gamma_{\beta, \epsilon}^{(2)}
\end{matrix}\right),
\intertext{where}
\left(\gamma_{\beta, \epsilon}^{(1)}\right)_{k, l} &= \frac{1}{2} \left[\tilde{\delta}_{k - l} \left(\left(\hat{X}_{\beta, \epsilon}\right)_{k} + \frac{1}{\left(\hat{X}_{\beta, \epsilon}\right)_{k}}\right) - \tilde{\delta}_{k + l} \left(\left(\hat{X}_{\beta, \epsilon}\right)_{k} - \frac{1}{\left(\hat{X}_{\beta, \epsilon}\right)_{k}}\right) \right],\\
\left(\gamma_{\beta, \epsilon}^{(2)}\right)_{k, l} &= \frac{1}{2} \left[\tilde{\delta}_{k - l} \left(\left(\hat{X}_{\beta, \epsilon}\right)_{k} + \frac{1}{\left(\hat{X}_{\beta, \epsilon}\right)_{k}}\right) + \tilde{\delta}_{k + l} \left(\left(\hat{X}_{\beta, \epsilon}\right)_{k} - \frac{1}{\left(\hat{X}_{\beta, \epsilon}\right)_{k}}\right) \right],\\
\left(\hat{X}_{\beta, \epsilon}\right)_{k} &= \sum_{j=0}^{N-1} e^{-2 \pi i \frac{k j}{N}} \left(X_{\beta, \epsilon}\right)_{j},
\intertext{and}
\tilde{\delta}_k &= \begin{cases}
1 &\text{if } k \text{ mod } N = 0,\\
0 &\text{otherwise}.
\end{cases}
\end{align}

Next, we trace out the momenta $k \notin A$ where
$A = \{1, \dots, \lfloor\frac{N-1}{2}\rfloor\}$,
i.e., we remove the negative momenta
and the momenta $k \in \{0, \frac{N}{2}\}$.
The resulting covariance matrix is diagonal:
\begin{align}
\label{eq:1d-covariance-matrix-in-momentum-space-after-taking-partial-trace}
 \mathcal{D}_A \mathcal{F} \gamma_{\beta, \epsilon} \mathcal{F}^t \mathcal{D}_A^t &= \bigoplus_{k=1}^{\lfloor \frac{N - 1}{2} \rfloor} \frac{1}{2} \left[\left(\hat{X}_{\beta}\right)_k + \frac{1}{\left(\hat{X}_{\beta}\right)_k}\right]
\oplus \bigoplus_{k=1}^{\lfloor \frac{N - 1}{2} \rfloor} \frac{1}{2} \left[\left(\hat{X}_{\beta}\right)_k + \frac{1}{\left(\hat{X}_{\beta}\right)_k}\right],
\end{align}
where we replaced $\left(\hat{X}_{\beta, \epsilon}\right)_{k}$ by
\begin{align}
\left(\hat{X}_{\beta}\right)_{k} &= \sum_{j=0}^{N-1} e^{-2 \pi i \frac{k j}{N}} \left(X_{\beta}\right)_{j}
\end{align}
since $\left(\hat{X}_{\beta}\right)_{k} = \left(\hat{X}_{\beta, \epsilon}\right)_{k}$ for $k \neq 0$.
In Eq.~\eqref{eq:1d-covariance-matrix-in-momentum-space-after-taking-partial-trace}, we
can directly read off the symplectic eigenvalues
of the reduced state's covariance matrix as
\begin{align}
\nu_k = \frac{1}{2} \left[\left(\hat{X}_{\beta}\right)_k + \frac{1}{\left(\hat{X}_{\beta}\right)_k}\right].
\end{align}
In a suitable basis, the entanglement Hamiltonian is thus given by $\sum_{k=1}^{\lfloor \frac{N - 1}{2} \rfloor} \tilde{\omega}_k b^\dagger_k b_k$, where $b_k$ and $b_k^\dagger$ are bosonic annihilation
and creation operators, and the entanglement energies are given by
\begin{align}
\tilde{\omega}_k &= \ln \left(\frac{\nu_k + 1}{\nu_k - 1}\right) = 2 \ln\left|\frac{\left(\hat{X}_{\beta}\right)_{k} + 1}{\left(\hat{X}_{\beta}\right)_{k} - 1}\right|.
\end{align}
Since we traced out the mode $k=0$, all entanglement energies are independent
of $\epsilon$.
\end{widetext}
\subsection{Entanglement spectrum on the cylinder}
\label{sec:appendix-entanglement-spec-cylinder-fourier-transform}
The entanglement cut of the cylinder made in Sec.~\ref{sec:psi-cont-2d-entanglement-spec}
preserves translational symmetry. Therefore, it is convenient
to express the eigenbasis of the entanglement Hamiltonian
in terms of Fourier modes as explained in the following.
The entanglement Hamiltonian is diagonal in the basis
that transforms the reduced state's covariance matrix
into Williamson's normal form.

For a cylinder of size $N_x \times N_y$ with $N_x$ even
and coordinates defined in Eq.~\eqref{eq:psi-cont-coordinates-cylinder}, we consider
the bipartition $A = \{1, \dots, \frac{N}{2}\}$, $B = \{\frac{N}{2} + 1, \dots, N\}$
corresponding to Fig.~\ref{fig:cylinder-cut}.
The Fourier transform of the reduced
state's covariance matrix is given by
\begin{align}
\mathcal{F}_y \gamma^{(A)}_{\beta, \epsilon} \mathcal{F}_y^\dagger &= \bigoplus_{k_y=0}^{N_y-1} \gamma^{(A)}_{k_y, \beta, \epsilon},
\end{align}
where $\gamma^{(A)}_{\beta, \epsilon} = \mathcal{D}_A \gamma_{\beta, \epsilon} \mathcal{D}_A^t$, and
\begin{align}
\mathcal{F}_y &= \left(\begin{matrix}
F_y & 0\\
0 & F_y^{*}
\end{matrix}\right)
\intertext{with}
\left(F_y\right)_{i_x k_y, j_x j_y} &= \frac{\delta_{i_x j_x}}{\sqrt{N_y}} e^{-\frac{2 \pi i}{N_y} k_y j_y}.
\end{align}
The $N_x \times N_x$ matrices
$\gamma^{(A)}_{k_y, \beta, \epsilon}$ are the blocks
of the reduced state's covariance matrix in momentum space.
We note that $\gamma^{(A)}_{k_y, \beta, \epsilon}$
is real
and $\gamma^{(A)}_{k_y, \beta, \epsilon} = \gamma^{(A)}_{-k_y, \beta, \epsilon}$.
This is a consequence of $X_{\beta, \epsilon}$
being symmetric under $i_y - j_y \to N_y - (i_y - j_y)$
for $y$ indices $i_y, j_y \in \{1, \dots, N_y\}$.

The matrix $\mathcal{F}_y$ is complex and does therefore
not define a real symplectic transformation. However,
we can construct a real matrix from it by combining the Fourier
modes of momentum $k_y$ and $-k_y$. After this
transformation, we have a description in terms of
positive momenta
$l_y \in \{0, \dots, \lfloor\frac{N_y}{2}\rfloor\}$
and an additional index $\sigma \in \{+, -\}$ for
$l_y \notin \{0, \frac{N_y}{2}\}$ and $\sigma = +$
for $l_y \in \{0, \frac{N_y}{2}\}$.
More precisely, we define the unitary matrix $T_y$ through
its action on a vector $c_{i_x k_y}$ in Fourier space
as
\begin{align}
\left(T_y c\right)_{i_x l_y \sigma} &= \begin{cases}
c_{i_x l_y}& \text{if } l_y \in \{0, \frac{N_y}{2}\},\\
\frac{1}{\sqrt{2}} \left(c_{i_x l_y} + c_{i_x, -l_y}\right)&
\text{if } l_y \notin  \{0, \frac{N_y}{2}\}\\& \text{and } \sigma = +,\\
\frac{1}{\sqrt{2}} \left(-i c_{i_x l_y} + i c_{i_x, -l_y}\right)&
\text{if } l_y \notin  \{0, \frac{N_y}{2}\}\\& \text{and } \sigma = -.
\end{cases}
\end{align}
Introducing
\begin{align}
\mathcal{T}_y &= \left(\begin{matrix}
T_y & 0\\
0 & T_y^{*}
\end{matrix}\right),
\end{align}
it follows that $\mathcal{T}_y \mathcal{F}_y$
is real and symplectic. Furthermore,
\begin{align}
\mathcal{T}_y \mathcal{F}_y \gamma^{(A)}_{\beta, \epsilon} \left(\mathcal{T}_y \mathcal{F}_y\right)^{t} &= \bigoplus_{l_y=0}^{\lfloor \frac{N_y}{2} \rfloor} \bigoplus_{\sigma}  \gamma^{(A)}_{l_y, \beta, \epsilon},
\end{align}
which follows from $\gamma^{(A)}_{l_y, \beta, \epsilon} = \gamma^{(A)}_{-l_y, \beta, \epsilon}$. Using
Williamson's decomposition~\cite{Williamson1936, DeGosson2006, Pirandola2009},
we next construct symplectic
matrices $R_{l_y, \beta, \epsilon}'$ to that
\begin{align}
&R_{l_y, \beta, \epsilon}' \gamma^{(A)}_{l_y, \beta, \epsilon} R_{l_y, \beta, \epsilon}'^t \\&\quad= \mathrm{diag}\left(\nu^{(1)}_{l_y, \beta, \epsilon}, \dots, \nu^{(\frac{N_x}{2})}_{l_y, \beta, \epsilon}, \nu^{(1)}_{l_y, \beta, \epsilon}, \dots, \nu^{(\frac{N_x}{2})}_{l_y, \beta, \epsilon}\right).\notag
\end{align}

The reduced state's covariance matrix $\gamma^{(A)}_{l_y, \beta, \epsilon}$ is thus brought
into Williamson's normal form through the transformation
\begin{align}
\left(\begin{matrix}
\bm{Q}'\\
\bm{P}'
\end{matrix}\right) &= R_{\beta, \epsilon}' \mathcal{T}_y \mathcal{F}_y \left(\begin{matrix}
\bm{Q}\\
\bm{P}
\end{matrix}\right),
\end{align}
where
\begin{align}
\label{eq:Rp-blocks}
R_{\beta, \epsilon}' &= \bigoplus_{l_y=0}^{\lfloor \frac{N_y}{2} \rfloor} \bigoplus_{\sigma} R_{l_y, \beta, \epsilon}'.
\end{align}
In terms of creation and annihilation operators,
this transformation is given by
\begin{align}
\label{eq:trafo-a-to-b-for-diagonalizing-gamma-half-cylinder}
\left(\begin{matrix}
\bm{b}\\
\bm{b}^\dagger
\end{matrix}\right) &= R_{\beta, \epsilon} \mathcal{T}_y \mathcal{F}_y \left(\begin{matrix}
\bm{a}\\
\bm{a}^\dagger
\end{matrix}\right),
\end{align}
where $R_{\beta, \epsilon} = \mathcal{U} R_{\beta, \epsilon}' \mathcal{U}^\dagger$,
the matrix $\mathcal{U}$ is defined as in
Eq.~\eqref{eq:psi-cont-u-for-transition-to-a},
and we used $\mathcal{U} \mathcal{T}_y \mathcal{F}_y \mathcal{U}^\dagger = \mathcal{T}_y \mathcal{F}_y$.

For the blocks with momenta $l_y \notin \{0, \frac{N_y}{2}\}$, the transformation
of Eq.~\eqref{eq:trafo-a-to-b-for-diagonalizing-gamma-half-cylinder} is given by
\begin{align}
\label{eq:appendix-R-for-a-to-b-diagonalizing-entanglement}
\left(\begin{matrix}
\bm{b}_{l_y, \sigma=+}\\
\bm{b}^\dagger_{l_y, \sigma=+}
\end{matrix}\right) &= R_{l_y, \beta, \epsilon} \frac{1}{\sqrt{2}}\left(\begin{matrix}
\tilde{\bm{a}}_{l_y} + \tilde{\bm{a}}_{-l_y}\\
\tilde{\bm{a}}_{l_y}^\dagger + \tilde{\bm{a}}_{-l_y}^\dagger\\
\end{matrix}\right),\\
\left(\begin{matrix}
\bm{b}_{l_y, \sigma=-}\\
\bm{b}^\dagger_{l_y, \sigma=-}
\end{matrix}\right) &= R_{l_y, \beta, \epsilon} \frac{1}{\sqrt{2}}\left(\begin{matrix}
-i \tilde{\bm{a}}_{l_y} + i \tilde{\bm{a}}_{-l_y}\\
i \tilde{\bm{a}}_{l_y}^\dagger -i \tilde{\bm{a}}_{-l_y}^\dagger\\
\end{matrix}\right),
\end{align}
where $\bm{b}_{l_y, \sigma}$ and $\tilde{\bm{a}}_{l_y}$
are vectors of length $\frac{N_x}{2}$ corresponding
to $i_x \in \{1, \dots, \frac{N_x}{2}\}$,
$R_{l_y, \beta, \epsilon}$ are the blocks of $R_{\beta, \epsilon}$ defined
analogously to Eq.~\eqref{eq:Rp-blocks}, and
\begin{align}
\tilde{a}_{i_x k_y} &= \frac{1}{\sqrt{N_y}} \sum_{j_y=1}^{N_y} e^{-\frac{2 \pi i}{N_y} j_y k_y} a_{i_x j_y}
\end{align}
are the Fourier transformed annihilation operators.

For $k_y \neq 0$, the blocks $\gamma^{(A)}_{\beta, \epsilon}$
and thus $R_{l_y, \beta, \epsilon}$ do not depend on $\epsilon$,
which follows from the definition of $X_{\beta, \epsilon}$
in Eq.~\eqref{eq:definition-of-X-beta-epsilon}. Thus,
Eq.~\eqref{eq:appendix-R-for-a-to-b-diagonalizing-entanglement} leads to
to Eq.~\eqref{eq:basis-transformation-a-to-b-for-cylinder-cut} in the main text,
where we suppressed
the dependence of $R_{l_y, \beta, \epsilon}$ on $\beta$
for better readability.

\section{Parent Hamiltonian}
\label{appendix:parent-Hamiltonian-psireal-cont-s}
The covariance matrix in $\psireal_{\beta, \epsilon}(\bm{s})$ is given
by
\begin{align}
\left(\begin{matrix}
X^{-1}_{\beta, \epsilon} & 0\\
0 & X_{\beta, \epsilon}
\end{matrix}\right),
\end{align}
cf. Eq.~\eqref{eq:gamma-cov-pure-state}.
The following Hamiltonians have a ground state with the same
covariance matrix and are thus parent Hamiltonians of $\psireal_{\beta, \epsilon}(\bm{s})$:
\begin{align}
\label{eq:parent-Hamiltonian-psi-tilde-beta-definition}
\mathcal{H}_{\eta} &= \frac{1}{2} \sum_{i, j=1}^{2N} \left(\begin{matrix}
\bm{Q}\\
\bm{P}
\end{matrix}\right)_i \left(H_{\eta}\right)_{i j}
\left(\begin{matrix}
\bm{Q}\\
\bm{P}
\end{matrix}\right)_j,
\intertext{where}
H_{\eta} &= \left(\begin{matrix}
X^{1+\eta}_{\beta, \epsilon}& 0\\
0&X^{-1+\eta}_{\beta, \epsilon}
\end{matrix}\right),
\end{align}
$\eta$ is a real parameter, and we used a  general result~\cite{Schuch2006}
about the relationship between a block-diagonal Hamiltonian
and the corresponding ground-state covariance matrix.

To diagonalize $\mathcal{H}_{\eta}$, we choose an orthonormal eigenbasis $V_{\epsilon}$ of $X_{\beta, \epsilon}$:
\begin{align}
V_{\epsilon}^t X_{\beta, \epsilon}V_{\epsilon} &= \chi_{\beta, \epsilon},
\intertext{where}
\chi_{\beta, \epsilon} &=  \mathrm{diag}\left(c^{(1)}_{\beta, \epsilon}, \dots, c^{(N)}_{\beta, \epsilon}\right).
\end{align}
[The eigenbasis $V_{\epsilon}$ is independent of $\beta$ since
$X_{\beta, \epsilon}$ depends on $\beta$ through a term proportional
to the identity, cf. Eq.~\eqref{eq:definition-of-X-beta-epsilon}.]
The symplectic matrix
\begin{align}
S_{\beta, \epsilon} &= \left(\begin{matrix}
X^{-\frac{1}{2}}_{\beta, \epsilon} V_{\epsilon}& 0\\
0 &  X^{\frac{1}{2}}_{\beta, \epsilon} V_{\epsilon}
\end{matrix}\right)
\end{align}
transforms $H_{\eta}$ into
\begin{align}
&S_{\beta, \epsilon}^t H_{\eta} S_{\beta, \epsilon}\\
&\quad= \mathrm{diag}\left[\left(c^{(1)}_{\beta, \epsilon}\right)^\eta, \dots,  \left(c^{(N)}_{\beta, \epsilon}\right)^\eta, \left(c^{(1)}_{\beta, \epsilon}\right)^\eta, \dots,  \left(c^{(N)}_{\beta, \epsilon}\right)^\eta\right].\notag
\end{align}
Thus, the symplectic eigenvalues of $H_{\eta}$ are given by $\left(c^{(k)}_{\beta, \epsilon}\right)^{\eta}$. Defining single-particle
energies as
\begin{align}
\omega_k = \left(c^{(k)}_{\beta, \epsilon}\right)^{\eta},
\end{align}
we thus find a parent Hamiltonian
$\sum_{k=1}^N \omega_k b^\dagger_k b_k$
of $\psireal_{\beta, \epsilon}(\bm{s})$, where the new creation
and annihilation operators $\bm{b}$ and $\bm{b}^\dagger$
are related to the original ones ($\bm{a}$ and $\bm{a}^\dagger$)
through
\begin{align}
\label{eq:b-and-b-dagger-from-a-and-a-dagger-parent-H}
\left(\begin{matrix}
\bm{b}\\
\bm{b}^\dagger
\end{matrix}\right) &= \frac{1}{2} \left(\begin{matrix}
\chi_{\beta, \epsilon}^{\frac{1}{2}} + \chi_{\beta, \epsilon}^{-\frac{1}{2}}
&
\chi_{\beta, \epsilon}^{\frac{1}{2}} - \chi_{\beta, \epsilon}^{-\frac{1}{2}}
\\
\chi_{\beta, \epsilon}^{\frac{1}{2}} - \chi_{\beta, \epsilon}^{-\frac{1}{2}}
&
\chi_{\beta, \epsilon}^{\frac{1}{2}} + \chi_{\beta, \epsilon}^{-\frac{1}{2}}
\end{matrix}\right)
\left(
\begin{matrix}
V_{\epsilon}^t & 0\\
0 & V_{\epsilon}^t
\end{matrix}
\right)
\left(\begin{matrix}
\bm{a}\\
\bm{a}^\dagger
\end{matrix}\right).
\end{align}
With $-i \frac{\partial \psireal}{\partial s_k}(\bm{s}) = i (X Q)_k \psireal(\bm{s})$,
it follows that the excited states with a single mode assume the form
\begin{align}
b^\dagger_k | \psireal_{\beta} \rangle &= \sqrt{2} \left(\chi_{\beta, \epsilon}^{\frac{1}{2}} V_{\epsilon}^t \bm{Q}\right)_k | \psireal_{\beta} \rangle.
\end{align}

The main text discusses the case $\eta = -1$, where the parent
Hamiltonian becomes
\begin{align}
\label{eq:parent-H-cont-spin-eta-eq-minus-1}
\mathcal{H}_{\eta=-1} &= \frac{1}{2} \left(\sum_{m=1}^N Q_m^2 + \sum_{m, n=1}^N C_{mn} P_m P_n\right)
\intertext{with}
C &= X_{\beta, \epsilon}^{-2}\big|_{\epsilon = 0} = \left[X_{\beta}^{-1} - \frac{X^{-1}_{\beta} \bm{e} \bm{e}^t X^{-1}_{\beta}}{\bm{e}^t X^{-1}_{\beta} \bm{e}}\right]^{2}.
\end{align}
We did numerical computations for a uniform lattice on the circle
and a square lattice on the cylinder
and
found that the matrix $C_{mn}$ decays with the distance between
the sites $m$ and $n$. At large distances,
the decay is consistent with a power law
on the circle and on the edge of the cylinder and with an exponential
in the bulk of the cylinder.

\section{Relation to lattice states obtained from conformal field theory}
\label{appendix:relation-to-previous-approximation}
We now relate the wave functions $\psireal_{\beta}(\bm{s})$
with $s_j$ continuous to
a class of lattice states $\psi_{\alpha}$ with
discrete spin-$\frac{1}{2}$ degrees of freedom constructed from CFT.
The latter are defined as
\begin{align}
\label{eq:def-of-psialpha}
|\psi_{\alpha}\rangle &= \sum_{s_1, \dots, s_N} \psi_{\alpha}(s_1, \dots, s_N)|s_1, \dots, s_N\rangle,\\
\psi_{\alpha}(s_1, \dots, s_N) &= \langle :e^{i \sqrt{\alpha} s_1\varphi(z_1)}: \dots :e^{i \sqrt{\alpha} s_N\varphi(z_N)}:\rangle\\
&= \delta_{\bm{s}} \prod_{m < n} (z_m - z_n)^{\alpha s_m s_n}
\end{align}
where $\delta_{\bm{s}} = 1$ if $s_1 + \dots + s_N = 0$ and $\delta_{\bm{s}} = 0$ otherwise, $\alpha > 0$ is a real parameter,  $s_i \in \{-1, 1\}$,
and $|s_1, \dots, s_N\rangle$ is the product of eigenstates
$|s_j\rangle$ of the spin-$z$ operator at lattice site $j$ ($t^z_j |s_j\rangle = \frac{1}{2} s_j | s_j \rangle$).

In a previous study~\cite{Herwerth2017}, we considered $zz$ correlations $C_{ij}$
in $\psi_{\alpha}$,
\begin{widetext}
\begin{align}
C_{ij} \equiv 4 \frac{\langle \psi_{\alpha}| t^z_i t^z_j | \psi_{\alpha}\rangle}{\langle \psi_{\alpha}| \psi_{\alpha}\rangle} &= \frac{\sum_{s_1, \dots, s_N} s_i s_j \delta_{\bm{s}} \prod_{m < n} |z_m - z_n|^{2 \alpha s_m s_n}}{\sum_{s_1, \dots, s_N} \delta_{\bm{s}}  \prod_{m < n} |z_m - z_n|^{2 \alpha s_m s_n}},
\end{align}
and approximated them for $i \neq j$ by a continuous integral of the form
\begin{align}
\label{eq:approximation-of-zz-correlations}
 C_{ij}^{(\mathrm{approx})} = \frac{\int d^N \bm{s} s_i s_j \delta\left(s_1 + \dots + s_N\right) e^{-\frac{1}{2} \bm{s}^2} \prod_{m < n} \left|\lambda \left(z_m - z_n\right)\right|^{2 \alpha s_m s_n}}{\int d^N \bm{s} \delta\left(s_1 + \dots + s_N\right)  e^{-\frac{1}{2} \bm{s}^2} \prod_{m < n} \left|\lambda (z_m - z_n)\right|^{2 \alpha s_m s_n}},
\end{align}
where $\lambda > 0$ is a scale parameter that we fixed
by requiring that the subleading
term of the approximation is minimal.

Rescaling $s_j \to s_j/\sqrt{\alpha}$ in $C^{(\mathrm{approx})}_{ij}$
and using $\prod_{m < n} \lambda^{s_m s_n} = e^{-\frac{1}{2} \bm{s}^2 \ln \lambda}$ for $s_1 + \dots + s_N = 0$,
we find
\begin{align}
C^{(\mathrm{approx})}_{ij} &= \frac{1}{\alpha}\frac{\int d^N \bm{s} s_i s_j \delta\left(s_1 + \dots + s_N\right) e^{-\frac{1}{2} \bm{s}^2 \left(\frac{1}{\alpha} + 2 \ln \lambda \right) } \prod_{m < n} \left|z_m - z_n\right|^{2 s_m s_n}}{\int d^N \bm{s} \delta\left(s_1 + \dots + s_N\right)  e^{-\frac{1}{2} \bm{s}^2 \left(\frac{1}{\alpha} + 2 \ln \lambda \right)} \prod_{m < n} \left|z_m - z_n\right|^{2 s_m s_n}}\\
&= \frac{1}{\alpha} \frac{\int d^N \bm{s} s_i s_j |\psireal_{\beta}(\bm{s})|^2}{\int d^N \bm{s} |\psireal_{\beta}(\bm{s})|^2},
\end{align}
where we identified $\beta + \beta_0 = \frac{1}{\alpha} + 2 \ln \lambda$. 
Thus, the approximation $C^{(\mathrm{approx})}_{ij}$ made in
Ref.~\onlinecite{Herwerth2017} corresponds to the spin-spin correlations
of the wave function $\psireal_{\beta}(\bm{s})$ up to a total factor of $\frac{1}{\alpha}$.
The same is true for the chiral state $\psichiral_{\beta}(\bm{s})$
of Sec.~\ref{sec:psi-cont-chiral-state}
since $\psireal_{\beta}(\bm{s})$ and $\psichiral_{\beta}(\bm{s})$
only differ by a phase and thus have the same spin-spin correlations.
\end{widetext}

\section{Chiral state}
\label{sec:psi-cont-chiral-state}
The chiral analog of $\psireal_{\beta}(\bm{s})$ has the form
\begin{align}
\label{eq:definition-psi-chiral-as-correlator}
\psichiral_{\beta}(\bm{s}) &= e^{-\frac{1}{4} \left(\beta + \beta_{0}\right) \bm{s}^2}  \langle :e^{i s_1 \varphi(z_1)}: \dots :e^{i s_N \varphi(z_N)}: \rangle\\
&=e^{-\frac{1}{4} \left(\beta + \beta_{0}\right) \bm{s}^2} \delta\left(s_1 + \dots + s_N\right)\notag \\
&\quad\times \prod_{m < n} \left(z_m - z_n\right)^{s_m s_n},\notag
\end{align}
where $\varphi(z)$ is the chiral part of the free boson
according to the decomposition $\varphi(z, \bar{z}) = \varphi(z) +  \bar{\varphi}(\bar{z})$.

On the cylinder introduced in Sec.~\ref{sec:cont-spin-states-cylinder}, the chiral wave function assumes
the form
\begin{align}
\label{eq:definition-psi-chiral-as-correlator-cylinder}
\psichiral_{\beta}(\bm{s}) &=  e^{-\frac{1}{4} \left(\beta + \beta_{0}\right) \bm{s}^2}  \langle :e^{i s_1 \varphi(w_1)}: \dots :e^{i s_N \varphi(w_N)}: \rangle\\
&= e^{-\frac{1}{4} \left(\beta + \beta_{0}\right) \bm{s}^2} \delta\left(s_1 + \dots + s_N\right)\notag \\
&\quad\times  \prod_{m < n}\left[2 \sinh\left(\frac{1}{2} (w_m - w_n)\right)\right]^{s_m s_n}\notag.
\end{align}

We now show that $\psireal_{\beta}(\bm{s})$
is equivalent to $\psichiral_{\beta}(\bm{s})$
for a uniform lattice in 1D with periodic boundary conditions.
Setting $N_x = 1$ in Eqs.~\eqref{eq:psi-cont-coordinates-cylinder}
and~\eqref{eq:definition-psi-chiral-as-correlator-cylinder}, we have
\begin{align}
\psichiral_{\beta}(\bm{s}) &=  \delta\left(s_1 + \dots + s_N\right) e^{-\frac{1}{4} (\beta + \beta_0) \bm{s}^2}\\
&\quad\quad\times \prod_{m < n}^{N} \left[2 i \sin\left(\frac{\pi}{N} (m - n)\right)\right]^{s_m s_n}\notag \\
&= e^{i \frac{\pi}{4} \bm{s}^2} \delta\left(s_1 + \dots + s_N\right) e^{-\frac{1}{4} (\beta + \beta_0) \bm{s}^2} \notag \\
&\quad\quad\times \prod_{m < n}^{N} \left[2 \sin\left(\frac{\pi}{N} (n - m)\right) \right]^{s_m s_n}\notag \\
&= e^{i \frac{\pi}{4} \bm{s}^2} \psireal_{\beta}(\bm{s})\notag,
\end{align}
where we used that $s_1 + \dots + s_N = 0$.
Up to the phase factor $e^{i \frac{\pi}{4} \bm{s}^2}$, the wave function
$\psichiral_{\beta}(\bm{s})$ thus coincides with
$\psireal_{\beta}(\bm{s})$.
Since $e^{i \frac{\pi}{4} \bm{s}^2}$ is a product of local phase factors,
it does not influence the spin-spin correlations and entanglement
entropies.
Furthermore, it can be transformed away through the symplectic
transformation
\begin{align}
\label{eq:symplectic-gauge-trafo-1d}
\left(\begin{matrix}
\bm{Q}\\
\bm{P}
\end{matrix}\right) &\to \left(\begin{matrix}
\mathbb{I}&0\\
-\frac{\pi}{2} \mathbb{I}&\mathbb{I}
\end{matrix}\right) \left(\begin{matrix}
\bm{Q}\\
\bm{P}
\end{matrix}\right).
\end{align}

For generic lattice configurations, however,
an important difference between the real
wave function $\psireal_{\beta}(\bm{s})$
and the chiral $\psichiral_{\beta}(\bm{s})$
is that the former is independent of the
lattice ordering while the latter is not.
More precisely, $\psireal_{\beta}(\bm{s})$
is symmetric under a simultaneous permutation $\sigma$
of both the spins $s_j$ and the positions $z_j$:
\begin{align}
\psireal_{\beta}^{(z_{\sigma(1)}, \dots, z_{\sigma(N)})}(s_{\sigma(1)}, \dots, s_{\sigma(N)}) &= \psireal_{\beta}^{(z_1, \dots, z_N)}(s_1, \dots, s_N),
\end{align}
where we explicitly wrote the parametric dependence of $\psireal_{\beta}(\bm{s})$ on
the lattice positions. This symmetry does, however,
not hold in the chiral case.
The permutation $\sigma(j) = N - j+1$, for example,
changes the wave function according to
\begin{align}
&\psichiral_{\beta}^{(z_{\sigma(1)}, \dots, z_{\sigma(N)})}(s_{\sigma(1)}, \dots, s_{\sigma(N)}) \\&\quad= 
\prod_{m < n} e^{\pi i s_m s_n \Phi_{n m}}\psichiral_{\beta}^{(z_1, \dots, z_N)}(s_1, \dots, s_N),\notag
\end{align}
where $\Phi_{m n} = \frac{1}{\pi} \left[\mathrm{arg} (z_m - z_n) - \mathrm{arg} (z_n - z_m)\right]$, and we explicitly wrote the parametric
dependence on $z_j$.
The phase $\prod_{m < n} e^{\pi i s_m s_n \Phi_{n m}}$
does not change the spin-spin correlations
in $\psichiral_{\beta}(\bm{s})$, but it influences entanglement properties
since it is in general not a product of single-particle phase factors.

\bibliography{refs}

\end{document}